\documentclass[aps,prd,twocolumn,a4paper,superscriptaddress,10pt,floatfix]{revtex4-2}
\usepackage{graphicx}
\usepackage{multirow}
\usepackage{latexsym}
\usepackage{hyperref}
\usepackage[british]{babel}
\usepackage{amsmath,amsmath}
\usepackage{color}
\usepackage[dvipsnames]{xcolor}
\usepackage{amsmath}
\begin{document}
\def\cjm#1{{\color{ForestGreen} \sf #1}}
\newcommand{\be}{\begin{equation}}
\newcommand{\ee}{\end{equation}}
\newcommand{\bq}{\begin{eqnarray}}
\newcommand{\eq}{\end{eqnarray}}

\title{Observational constraints on vector-like dark energy}
\author{C. S. C. M. Coelho}
\email{carolinacastrocoelho@gmail.com}
\affiliation{Escola Secund\'aria Carolina Micha{\"e}lis, Rua Infanta D. Maria 159–175, 4050-350 Porto, Portugal}
\author{A.-L. Y. Gschrey}
\email{An.Gschrey@campus.lmu.de}
\affiliation{Department of Physics, Ludwig-Maximilians-Universität M{\"u}nchen,  Geschwister-Scholl-Platz 1, D-80539 M{\"u}nchen, Germany}
\author{C. J. A. P. Martins}
\email{Carlos.Martins@astro.up.pt}
\affiliation{Centro de Astrof\'{\i}sica da Universidade do Porto, Rua das Estrelas, 4150-762 Porto, Portugal}
\affiliation{Instituto de Astrof\'{\i}sica e Ci\^encias do Espa\c co, Universidade do Porto, Rua das Estrelas, 4150-762 Porto, Portugal}

\date{\today}
\begin{abstract}
The canonical cosmological model to explain the recent acceleration of the universe relies on a cosmological constant, and most dynamical dark energy and modified gravity model alternatives are based on scalar fields. Still, further alternatives are possible. One of these involves vector fields: under certain conditions, they can lead to accelerating universes while preserving large-scale homogeneity and isotropy. We report quantitative observational constraints on a model previously proposed by Armend\'ariz-Pic\'on and known as the cosmic triad. We consider several subclasses of the model, which generically is a parametric extension of the canonical $\Lambda$CDM model, as well as two possible choices of the triad's potential. Our analysis shows that any deviations from this limit are constrained to be small. In particular the preferred present-day values of the matter density and the dark energy equation of state are fully consistent with those obtained, for the same datasets, in flat $\Lambda$CDM and $w_0$CDM. The constraints mildly depend on the priors on the dark energy equation of state, specifically on whether phantom values thereof are allowed, while the choice of potential does not play a significant role since any such potential is constrained to be relatively flat.
\end{abstract}
\maketitle
\section{\label{intro}Introduction} 

Characterizing the physical mechanism responsible for the observed acceleration of the recent universe is the primary challenge of contemporary fundamental cosmology. A cosmological constant is the simplest explanation consistent with observations, leading to the canonical $\Lambda$CDM paradigm, but the observationally inferred value is in violent disagreement with particle physics expectations. It is therefore of interest to carry out consistency tests of the canonical scenario, while exploring alternative scenarios based on different physical mechanisms. Most such alternatives, usually classified under the broad categories of dynamical dark energy and modified gravity, rely on one or more scalar fields. These are the simplest possible dynamical degrees of freedom, and are endowed with the further advantage that they are known to be among nature's fundamental building blocks, cf. the Higgs field \cite{ATLAS,CMS}.

Still, further alternatives may be envisaged. A vector-like dark energy scenario was first considered in detail in \cite{Paper1}, and subsequently (with a different assumption for the vector potential) in \cite{Paper3}, while an alternative model of the same broad class but not including any potential was considered in \cite{Paper2a,Paper2b}. We will refer to them as the triad and dyad models respectively. These works argued on general grounds that such models may in principle be observationally viable, and presented some preliminary constraints on them (under various simplifying assumptions), but to our knowledge there is so far no quantitative comparison of these models with modern cosmological data. In what follows we report on one such analysis for the triad model, focusing on low-redshift background cosmology data. As will be shown, this is sufficient to tightly constrain it. We rely on the same analysis methodology to also briefly comment on the dyad model in an appendix. Other generic aspects of vector fields in cosmology have been addressed in \cite{Himmetoglu,Esposito,Referee1,Tasinato,Referee2,Referee3,Gomez}.

Superficially vector fields are not promising, since a single cosmological vector field would violate isotropy. However, a way to circumvent this was pointed in \cite{Paper1}: a cosmic triad, comprising three identical vector fields pointing in mutually orthogonal three-dimensional space directions can preserve isotropy (at least at the background gravitational level) and lead to late-time accelerating solutions which, at least in some circumstances, have a well-defined $\Lambda$CDM limit. In fact, depending on model parameters, the cosmic triad's equation of state can have either a canonical ($w\ge-1$) or a phantom behavior ($w<-1$). The alternative dyad model approach \cite{Paper2a} has various similarities with the triad model, but also significant differences.

We start in Sect. \ref{models} by briefly introducing the triad model and discussing representative cosmological solutions for the vector field. In Sect. \ref{methods} we outline our numerical implementation of this model, which enables its quantitative comparison, through standard statistical methods, with two canonical low-redshift background cosmology data sets, which we also briefly describe. We then report our results, first for some particular cases in Sect. \ref{result1}, and then for generic triads in Sect. \ref{result2}. Finally, we present our conclusions in Sect. \ref{conc}, and make a brief comparison to the dyad case in Appendix \ref{app1}. Throughout this work we use units with $c=1$.

\section{\label{models}Triad dark energy models} 

Our starting point are the Einstein and vector field equations for this class of models. In what follows we will not derive them in detail, but only state the main assumptions behind them, pointing the reader to \cite{Paper1} for detailed derivations. Therein one assumes a set of three self-interacting vector fields (more rigorously, they are one-forms), including a self-interaction term which, in general, would explicitly break gauge invariance. Moreover, one assumes that the matter Lagrangian does not depend on the cosmic triad. In this case, and in a homogeneous and isotropic Friedmann-Lema\^{\i}tre-Robertson-Walker universe, the Einstein equations take the usual mathematical form
\bq
H^2&=&\frac{\kappa^2}{3}\rho\\
\frac{\ddot a}{a}&=&-\frac{\kappa^2}{6}(\rho+3p)\,,
\eq
where $\kappa^2=8\pi G$ and dots denote derivatives with respect to physical time. In addition to standard components, the cosmic triad, denoted $A$, has contributions to the density and pressure terms of the form
\bq
\label{defr}
\rho_A&=&\frac{3}{2}({\dot A}+HA)^2+3V(A^2)\\
p_A&=&\frac{1}{2}({\dot A}+HA)^2-3V(A^2)+2A^2\frac{dV}{dA^2}\,.
\label{defp}
\eq
Note that it is assumed that the self-interaction potential has the form $V=V(A^2)$. Clearly in this model a potential is necessary for acceleration, since in its absence the triad's equation of state is always that of radiation, $w_A=1/3$. Additionally there is a Proca-like equation for the cosmic triad
\be\label{proca1}
{\ddot A}+3H{\dot A}+\left(H^2+\frac{\ddot a}{a}\right)A+\frac{dV}{dA}=0\,.
\ee
Inspection of these equations immediately reveals that the choice $(A=0,V=const.)$ corresponds to $\Lambda$CDM. Moreover, it is also apparent that a potential such that $dV/dA^2<0$ could lead, if the kinetic term is suitably small, to a phantom equation of state. The previous studies of \cite{Paper1} and \cite{Paper3} have assumed the following power law and exponential potentials, respectively,
\bq
\label{potential1}
V_1(A^2)&=&M^4\left(\frac{A^2}{M^2}\right)^{-n}\\
V_2(A^2)&=&M^4\exp{\left(-\kappa^2\lambda A^2\right)}\,,
\label{potential2}
\eq
which satisfy the negative slope condition stated in the previous paragraph for $n>0$ and $\lambda>0$ respectively. Thus in these models the phantom behavior stems from the shape of the potential, rather than from a kinetic term with the wrong sign.

It is interesting to exhibit the possible scaling behaviors of the vector fields, specifically proportional to $t^\alpha$, for generic power-law expanding universes, $a(t)\propto t^\beta$, with $0\le\beta<1$. We start with the case of a vanishing or constant potential, in which case there are two scaling solutions
\bq\label{conformal1}
A&\propto& \frac{1}{a}\\
A&\propto& \frac{t}{a^2}\,;
\label{conformal2}
\eq
the first is always a decaying triad model, while the second is a constant in the radiation era ($\beta=1/2$), and a growing or decaying model for slower and faster expansion rates respectively---the latter including the matter era. Note that a constant solution is indeed allowed in the radiation era, since in that case the Ricci scalar vanishes.

Curiously, if one retains the assumption of a power law scale factor, which physically corresponds to assuming a universe with a single component (e,g, radiation or matter), with the vector component having a negligible contribution to the Friedmann equation, the full Proca equation, including the potential given by Eq. (\ref{potential1}) always has a growing mode solution for general power law scale factors, given by \cite{Paper1}
\be\label{nobeta}
A(t)=\left(\frac{2n(1+n)^2M^{2(2+n)}}{\beta(2\beta-1)(1+n)^2+3\beta(1+n)-n}\right)^{\frac{1}{2+2n}} t^{\frac{1}{1+n}}\,,
\ee
which depends on the slope of the potential through its power $n$ but not the expansion rate $\beta$ except for the dependence of the constant pre-factor of $A$. This corresponds to an energy density evolving as $\rho_A\propto t^{-\frac{2n}{1+n}}$. For all values of $n\ge-1$ the decay is slower than the standard $\rho\propto t^{-2}$, so the above solution will necessarily be a transient one. No analogous solution exists for the alternative potential of Eq. (\ref{potential2}).

\begin{figure}
  \begin{center}
    \includegraphics[width=\columnwidth]{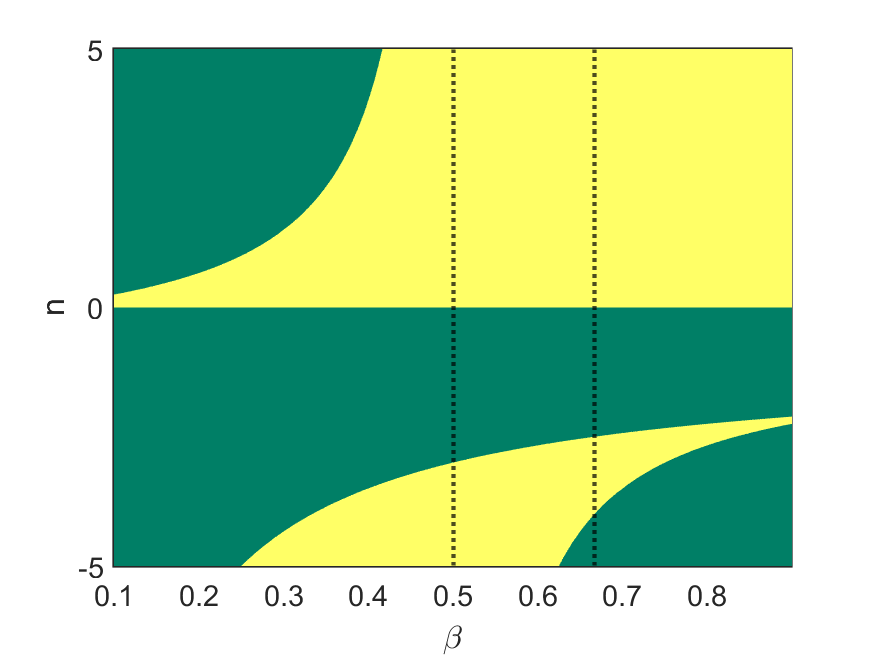}
    \caption{Domain of validity of Eq. (\ref{nobeta}), as a function of the cosmological expansion rate $\beta$ and the potential power law $n$. The solution is valid in the light-shaded regions. The vertical dotted lines identify the expansion rates corresponding to the radiation and matter eras.}
    \label{fig01}
  \end{center}
\end{figure}

Naturally, for this solution to exist the term inside the pre-factor brackets must be positive, which restricts the combinations of the parameters $(\beta,n)$ to the light-shaded regions in Fig. \ref{fig01}. In both the radiation and matter eras all plausible choices of $n>0$ are nominally allowed, although some negative values would be too. This suggests a choice of prior $n\ge0$, which we will adopt in some of the analysis which follows.

\section{\label{methods}Statistical analysis and numerical implementation} 

We use a standard grid-based statistical likelihood analysis, as described for example in \cite{Verde}. The likelihood function can be symbolically written
\be
{\cal L}(p)\propto\exp{\left(-\frac{1}{2}\chi^2(p)\right)}\,,
\ee
where $p$ denotes the free parameters of the model being studied---in our case, there will be up to four free parameters. For a generic redshift-dependent quantity $O(z)$, the chi-square function is
\be
\chi^2(p)=\sum_{i,j}\left(O_{d,i}-O_{m,i}(p)\right)C_{ij}^{-1}\left(O_{d,j}-O_{m,j}(p)\right)\,,
\ee
where the $d$ and $m$ subscripts denote the data and model respectively, and $C$ is the covariance matrix of the dataset.

Our analysis is limited to background cosmology observables, the main reason being that in these vector-based models linear cosmological perturbation theory is complicated by the presence of anisotropic stresses and, more significantly, by the absence of the usual decoupling of scalar, vector and tensor perturbations from each other \cite{Paper1}. Addressing these issues is beyond the scope of this work. We will also restrict ourselves to flat universes, setting the spatial curvature parameter to zero, in agreement with most contemporary cosmological observations \cite{Planck}.

We therefore use two canonical low-redshift background cosmology datasets. The Pantheon compilation \cite{Scolnic,Riess} contains 1048 Type Ia supernovae, compressed into 6 correlated measurements at different redshift bins in the range $0.07<z<1.5$. (It also relies on 15 Type Ia supernovae from two Hubble Space Telescope Multi-Cycle Treasury programs, and it assumes a spatially flat universe.) The compression methodology and validation are detailed in Section 3 of \cite{Riess}. We also use the compilation of 38 Hubble parameter measurements of Farooq {\it et al.} \cite{Farooq}. This is a more heterogeneous set, containing both cosmic chronometers and baryon acoustic oscillation data. Together, the two cosmological data sets contain measurements up to redshift $z\sim2.36$.

The Hubble constant is always analytically marginalized \cite{Homarg}: we trivially write $H(z)=H_0 E(z)$, whence $H_0$ is purely a multiplicative constant, and can be analytically integrated in the likelihood. Defining
\bq
X(p)&=&\sum_{i}\frac{E_{m,i}^2(p)}{\sigma^2_i}\\
Y(p)&=&\sum_{i}\frac{E_{m,i}(p) H_{d,i}}{\sigma^2_i}\\
Z(p)&=&\sum_{i}\frac{H_{d,i}^2}{\sigma^2_i}
\eq
where the $\sigma_i$ are the uncertainties in observed values of the Hubble parameter, the chi-square  with the $H_0$ parameter marginalized is given by
\be
\chi^2(q)=Z(q)-\frac{Y^2(q)}{X(q)}+\ln{X(q)}-2\ln{\left[1+Erf{\left(\frac{Y(q)}{\sqrt{2X(q)}}\right)}\right]}
\ee
where $Erf$ is the Gauss error function and $\ln$ is the natural logarithm. Therefore our results do not depend on possible choices of the Hubble constant, and in particular they are insensitive to the so-called Hubble tension. Unless otherwise is stated, we use uniform priors for the model parameters, in the plotted ranges. We have tested that these prior range choices do not significantly impact our results.

For the statistical analysis we need to simultaneously solve the Friedmann and Proca equations, which can more conveniently be done with dimensionless variables. It is also useful to express the Proca equation as two first-order differential equations. Defining
\be
B=\dot{A}+HA\,,
\ee
Eq. (\ref{proca1}) becomes
\bq
\label{step1}
(1+z)\frac{dA}{dz}&=&A-\frac{B}{H}\\
(1+z)\frac{dB}{dz}&=&2B+\frac{1}{H}\frac{dV}{dA}\,,
\label{step2}
\eq
which we have now expressed as a function of redshift. For the first of these equations it is clearly advantageous to define
\bq
A(z)&=&A_0 f(z)\\
B(z)&=&B_0 g(z)\,,
\eq
where the subscript $0$ denotes present-day values and we can define
\be
r\equiv \frac{B_0}{H_0A_0}=1+\frac{\dot f_0}{H_0}\,,
\ee
which generically is one of the model's free parameters. Note that the case of vanishing speed of the $A$ field corresponds to $r=1$, not $r=0$. Then Eq. (\ref{step1}) becomes
\be\label{base0}
(1+z)\frac{df(z)}{dz}=f(z)-r\frac{g(z)}{E(z)}\,;
\ee
note that if $r=0$ the solution is $f(z)=(1+z)$ and therefore $A\propto 1/a$, which is is one of the solutions discussed in the previous section.

For Eq. (\ref{step2}), we start by re-writing the potential of Eq. (\ref{potential1}) as $V_1=V_0(A/A_0)^{-2n}$. Using the definitions of the triad's density and pressure, cf. Eqs. (\ref{defr},\ref{defp}), we can obtain its present-day equation of state
\be
\label{middle1}
w_0=\frac{\frac{1}{2}B_0^2-(3+2n)V_0}{\frac{3}{2}B_0^2+3V_0}\,,
\ee
which we can invert to yield
\be
\label{middle2}
V_0=\frac{\frac{1}{2}(1-3w_0)B_0^2}{3(1+w_0)+2n}\,;
\ee
note that for $w_0=1/3$ we get $V_0=0$ (as we should, given the discussion in the previous section), while for $w_0=-1$ we get $V_0=B_0^2/n$. Then Eq. (\ref{step2}) becomes
\be\label{bv1}
(1+z)\frac{dg(z)}{dz}=2g(z)-\frac{(1-3w_0)nr}{2n+3(1+w_0)}\frac{1}{E(z)f^{2n+1}(z)}\,.
\ee
Again we observe the expected behavior: if $n=0$ or $w_0=1/3$ we have $g(z)=(1+z)^2=a^{-2}$; the same happens of $r=0$, but then we also have $B_0=0$.

Finally we can use these parameters to write the dimensionless Friedmann equation. Assuming a spatially flat universe containing matter and the vector field (neglecting radiation, since we are only concerned with low redshifts) and defining
\bq
\Omega_m &=& \frac{\kappa^2\rho_0}{3H_0^2}\\
\Omega_A &=& \frac{\kappa^2 B_0^2}{2H_0^2}\,, \label{defoma}
\eq
with the caveat that the latter only applies provided $B_0^2>0$, we have
\be
E^2=\Omega_m(1+z)^3+\Omega_A\left(g^2+\frac{(1-3w_0)f^{-2n}}{2n+3(1+w_0)}\right)\,,
\ee
but since by definition $E(0)=1$, $\Omega_m$ and $\Omega_A$ are not independent, and we can equivalently write
\be\label{friedv1}
E^2=\Omega_m(1+z)^3+\frac{(1-\Omega_m)}{2n+4}\left[(2n+3(1+w_0))g^2+\frac{1-3w_0}{f^{2n}}\right]\,,
\ee
from which it is clear that the $\Lambda$CDM limit obtains for the case $(n=0,w_0=-1)$. Thus we see that in general this model has a four-dimensional parameter space, $(\Omega_m,w_0,n,r)$. Note that $r$ does not appear explicitly in the Friedmann equation, but does affect it indirectly (and therefore can be constrained by background cosmology data) since it impacts the evolution of $f$ and $g$.

This process can be repeated for the alternative triad model potential, given by Eq. (\ref{potential2}) which we re-write as $V_2=V_0\exp[-\kappa^2\lambda(A^2-A_0^2)]$. The analogous expressions for $w_0$ and $V_0$ retain the same form as Eqs. (\ref{middle1},\ref{middle2}) with the trivial substitution $n\longrightarrow \lambda\kappa^2A_0^2\equiv m$, which we henceforth denote as $m$. Naturally Eq. (\ref{base0}) is unchanged while Eq. (\ref{bv1}) becomes
\be\label{bv2}
(1+z)\frac{dg}{dz}=2g-\frac{(1-3w_0)mr}{2m+3(1+w_0)}\frac{fe^{-m(f^2-1)}}{E(z)}\,.
\ee
The definition of $\Omega_A$ is also unchanged, and the Friedmann equation becomes
\be
E^2=\Omega_m(1+z)^3+\Omega_A\left(g^2+\frac{(1-3w_0)e^{-m(f^2-1)}}{2m+3(1+w_0)}\right)\,,
\ee
or equivalently
\be\label{friedv2}
E^2=\Omega_m(1+z)^3+\frac{(1-\Omega_m)}{2m+4}\left[(2m+3(1+w_0))g^2+\frac{1-3w_0}{e^{m(f^2-1)}}\right]\,,
\ee
whose $\Lambda$CDM limit still corresponds to the case $(m=0,w_0=-1)$. As expected, the different choice of potential is only explicitly manifest in a single model parameter ($n$ or $m$), with all the other parameters being the same.

\begin{figure*}
  \begin{center}
    \includegraphics[width=\columnwidth]{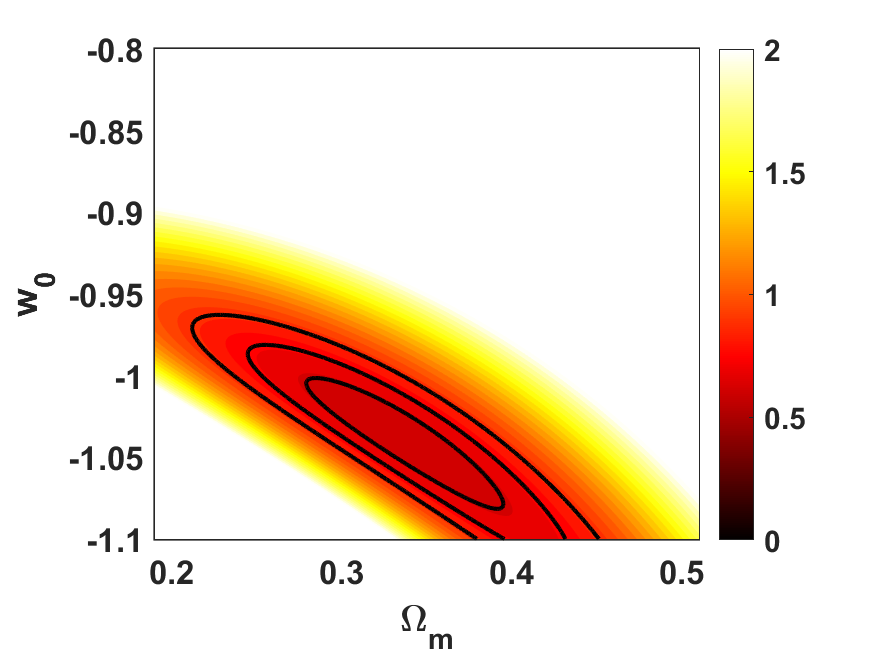}
    \includegraphics[width=\columnwidth]{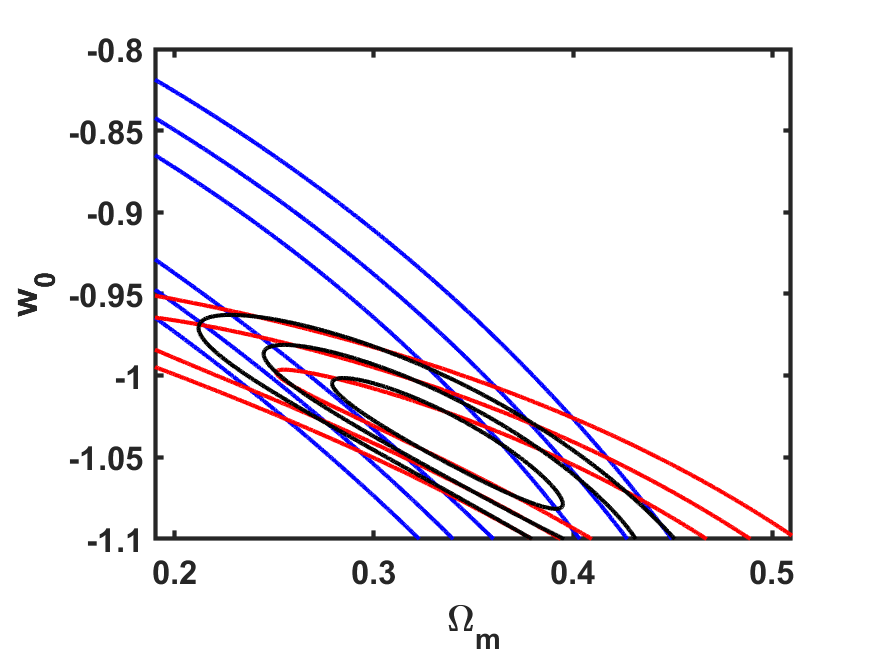}
    \includegraphics[width=\columnwidth]{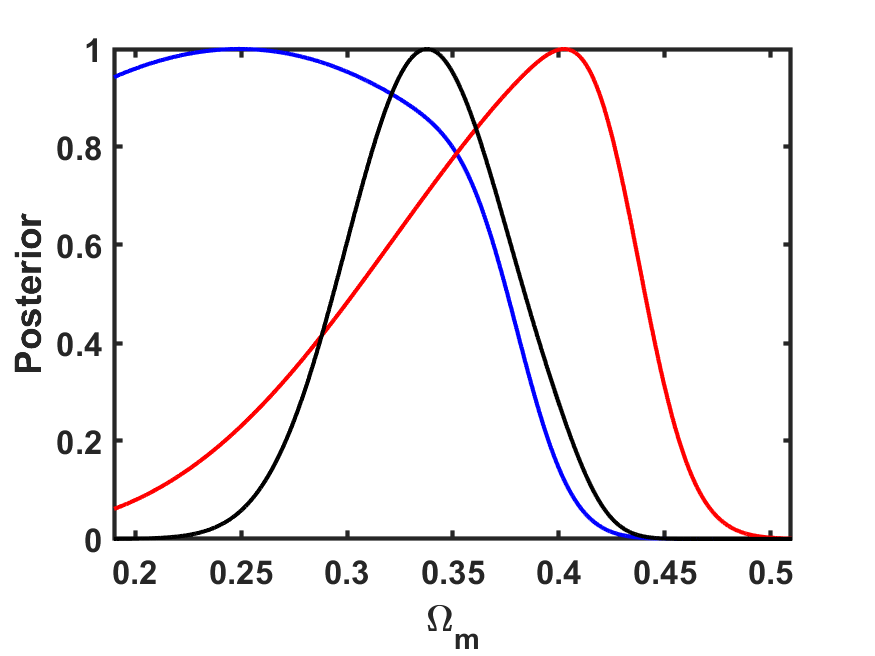}
    \includegraphics[width=\columnwidth]{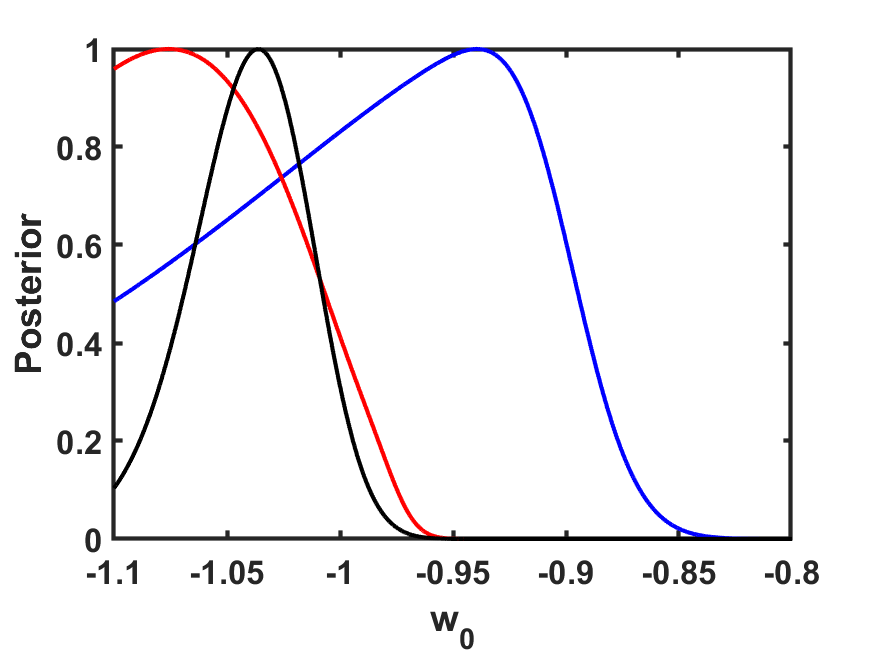}
    \caption{Constraints on the constant potential triad model. The top panels depict the constraints on the two-dimensional $\Omega_m$--$w_0$ plane: one, two and three sigma confidence levels are shown, with the colormap corresponding to the reduced chi-square. The bottom panels show the one-dimensional posteriors for the two parameters. Blue, red and black lines, correspond to the supernova data, the Hubble parameter data, and their combination, respectively.}
    \label{fig02}
  \end{center}
\end{figure*}

In the following two sections we present detailed constraints on the triad model, both for its full parameter space and for some particular cases thereof. Our statistical analysis uses the exact analytic solutions in the particular cases in which these exist (in which cases we provide them in the text), while in the other cases the exact Friedmann and Proca equations are solved using standard numerical integration methods.

\section{\label{result1}Constraints for specific triads}

Before embarking on a discussion of the constraints of the generic triad model, which we will do in the following section, we will consider the constraints for particular cases thereof, with the goal of developing a better understanding of its overall behavior. As a trivial starting point, we begin with the triad model without the potential. Setting $V=0$ implies $g(z)=(1+z)^2$ as already discussed, so the model's surviving equations are
\bq
(1+z)\frac{df}{dz}&=&f-r\frac{(1+z)^2}{E(z)}\\
E^2&=&\Omega_m(1+z)^3+(1-\Omega_m)(1+z)^4\,.
\eq
Moreover, one can actually integrate the first equation to obtain
\be
f(z)=\frac{2rE(z)}{\Omega_m(1+z)}+ (1+z)\left(1- \frac{2r}{\Omega_m}\right)\,.
\ee
In this limit the model includes matter plus a radiation-like component but no acceleration mechanism, and evidently it will not fit the data.

It is also interesting to consider the case of the triad with constant potential. In that case we still have $g(z)=(1+z)^2$ and the equation for $f$ remains the same, but setting $n=m=0$ leads to the following Friedmann equation
\be
E^2=\Omega_m(1+z)^3+\frac{(1-\Omega_m)}{4}\left[3(1+w_0)(1+z)^4+(1-3w_0)\right]\,,
\ee
which is obviously the same for both potentials. It is still the case that $r$ does not impact the Friedmann equation. Note that as long as these models are envisaged as phenomenological low-redshift ones, phantom equations of state ($w_0<-1$) are in principle allowed, but such a deviation towards the phantom side cannot be too large, since otherwise the radiation-like term would imply that the right-hand side of the Friedmann equation would become negative, and therefore unphysical.

\begin{figure*}
  \begin{center}
    \includegraphics[width=\columnwidth]{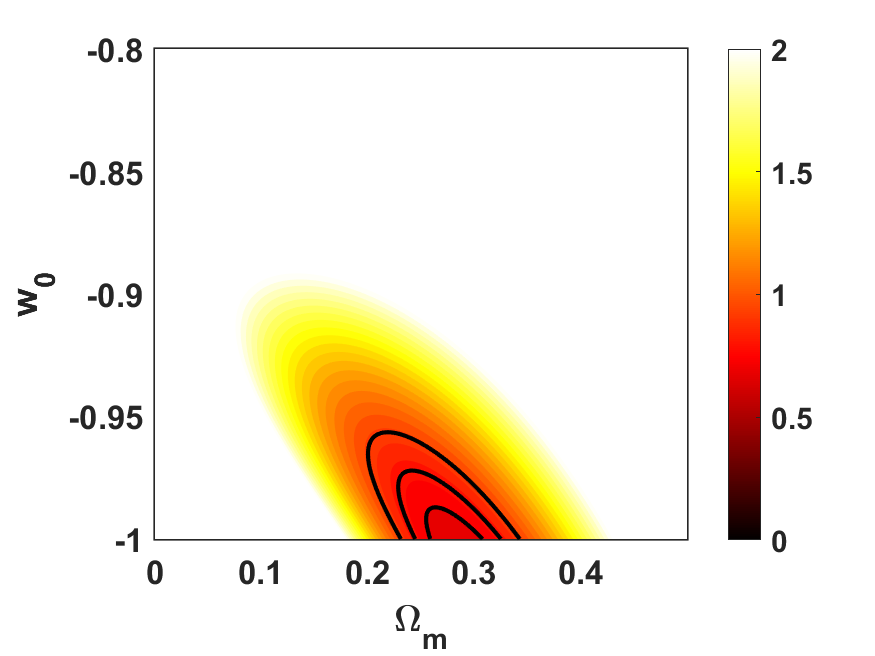}
    \includegraphics[width=\columnwidth]{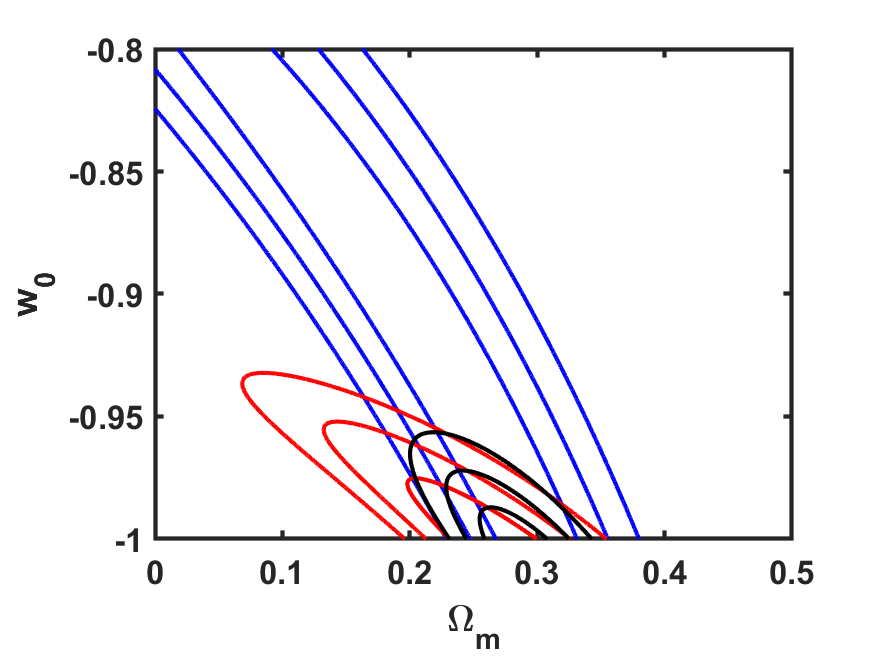}
    \includegraphics[width=\columnwidth]{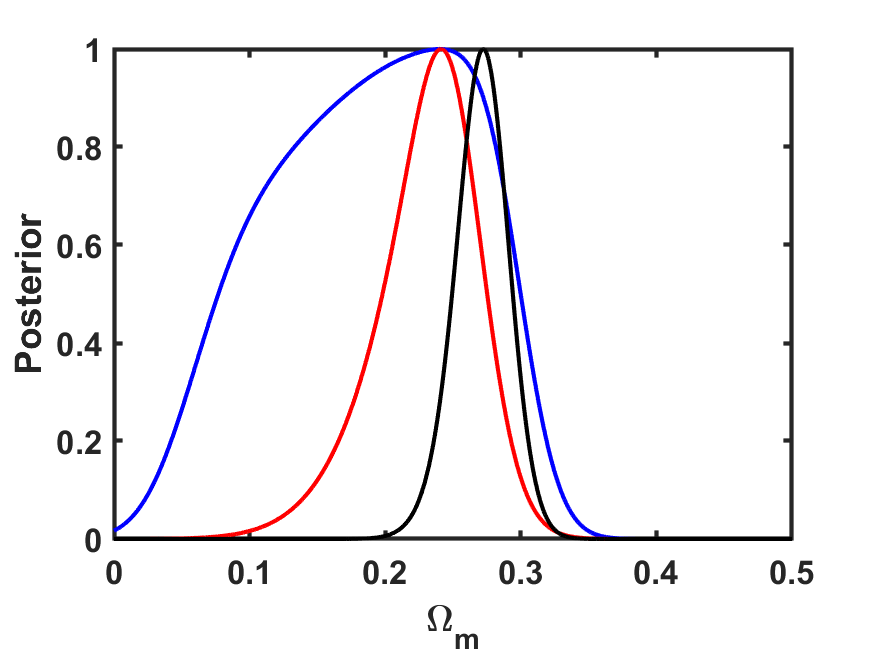}
    \includegraphics[width=\columnwidth]{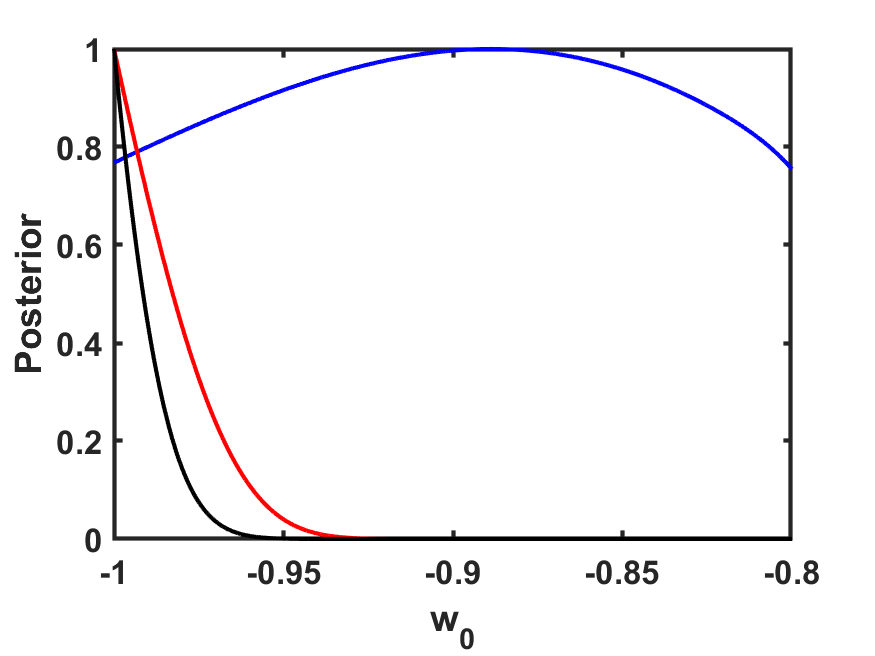}
    \caption{Same as Fig. \ref{fig02}, excluding the possibility of phantom equations of state. Note that the $\Omega_m$ range is also changed.}
    \label{fig03}
  \end{center}
\end{figure*}

Figure \ref{fig02} shows the constraints on this model, confirming that a $\Lambda$CDM-like model is preferred. As expected the two parameters are anticorrelated but the degeneracy directions are different for the supernova and Hubble parameter datasets, so their combination partially breaks the degeneracy and leads to the one-sigma constraints
\bq
\Omega_m&=&0.34\pm0.04\\
w_0&=&-1.04\pm0.03\,.
\eq
It is interesting to compare these constraints with those obtained, for the same datasets, for the standard flat $w_0$CDM model, for which
\be
E^2=\Omega_m(1+z)^3+(1-\Omega_m)(1+z)^{3(1+w_0)}\,.
\ee
In this case, \cite{Fernandes} reports
\bq\label{w0cdmm}
\Omega_m&=&0.27\pm0.02\\
w_0&=&-0.92\pm0.06\,;\label{w0cdmw}
\eq
therefore in the constant-potential triad the preferred value of the matter density is increased with respect to the $w_0$CDM case, and the preferred value of the dark energy equation of state is correspondingly decreased.

Due to the aforementioned parameter degeneracies, the constraints derived for the individual datasets will be somewhat dependent on the choice of priors, but this is much less so for the combined constraints. These dependencies are illustrated in Fig. \ref{fig03}, which shows analogous constraints in the case where phantom equations of state are not allowed. In this case we find
\bq
\Omega_m&=&0.27\pm0.02\\
w_0&<&-0.98\,,
\eq
where a two-sigma upper limit is given for $w_0$. Notice that the preferred value of the matter density is now smaller (as expected) and actually agrees with the preferred value for $w_0$CDM.

\begin{figure*}
  \begin{center}
    \includegraphics[width=\columnwidth]{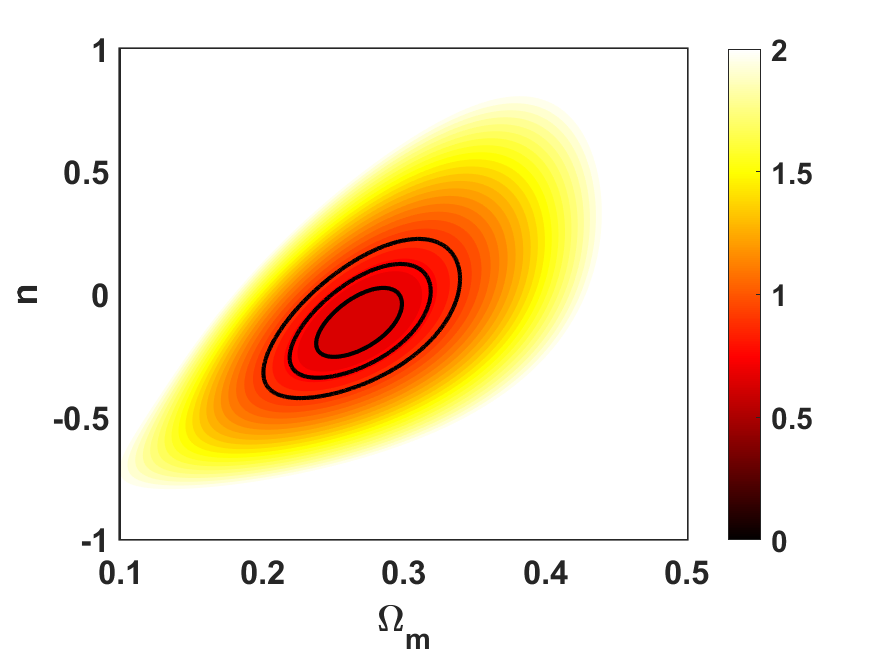}
    \includegraphics[width=\columnwidth]{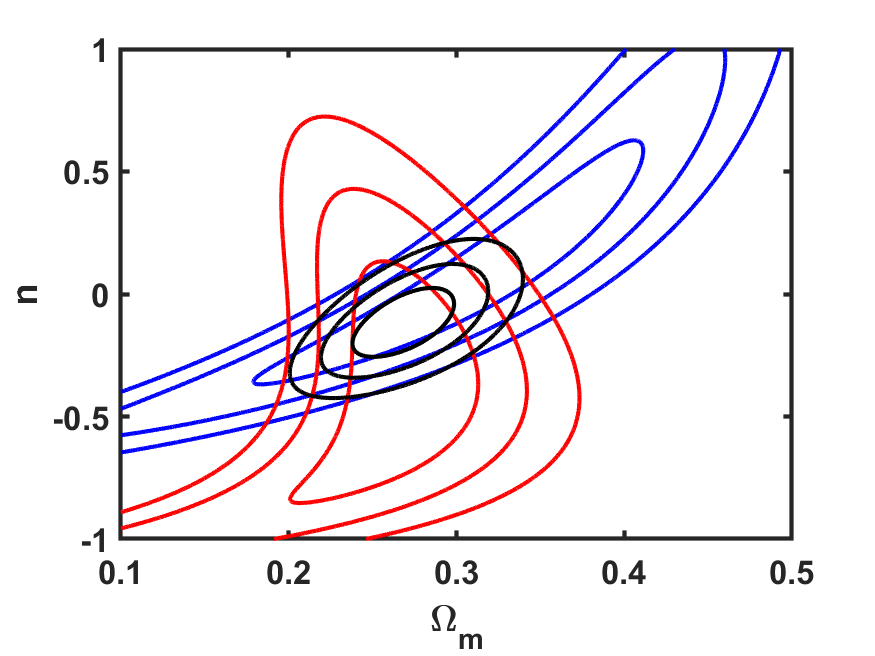}
    \includegraphics[width=\columnwidth]{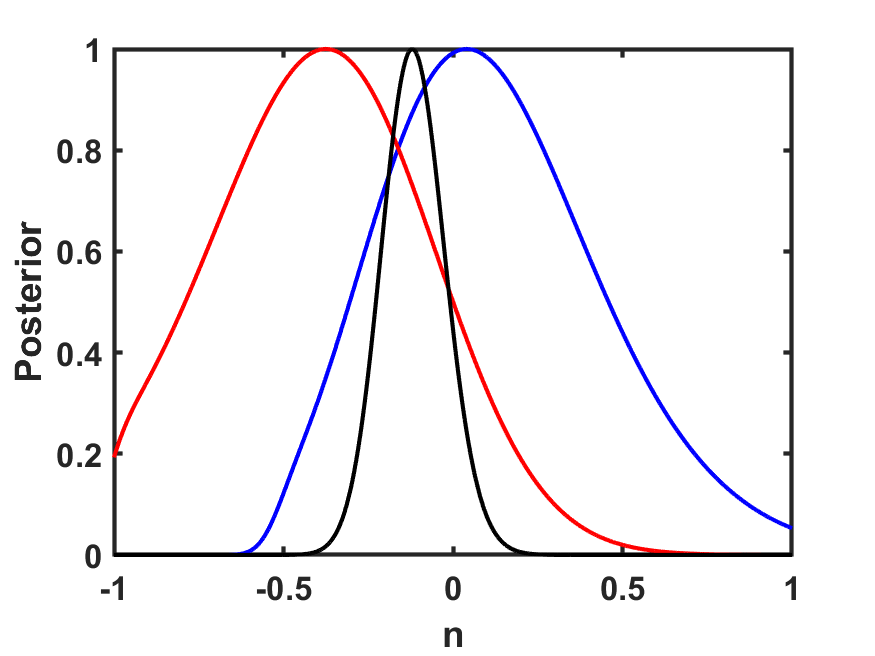}
    \includegraphics[width=\columnwidth]{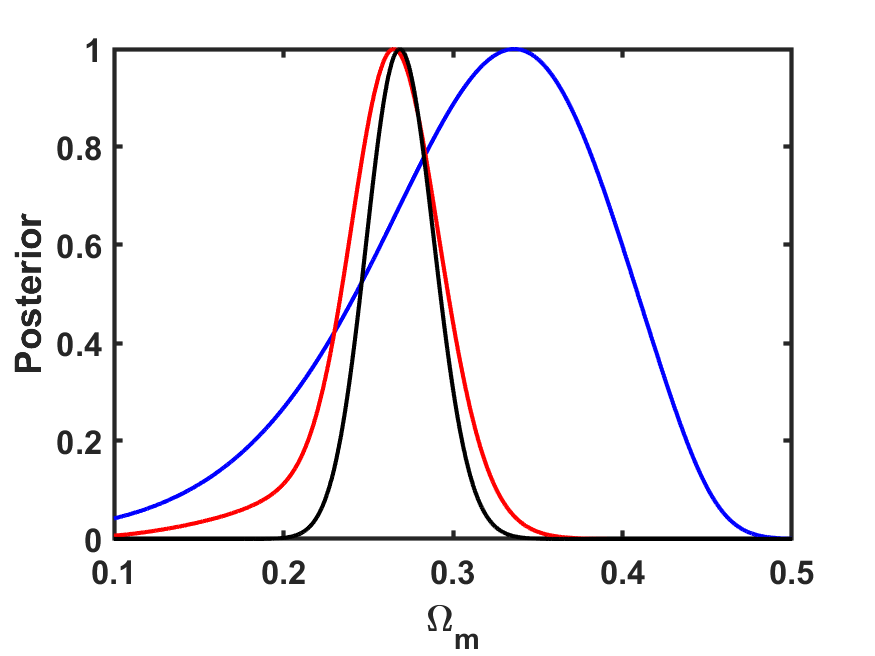}
    \caption{Constraints on the $r=0$ triad model. The top panels depict the constraints on the two-dimensional $\Omega_m$--$w_0$ plane: one, two and three sigma confidence levels are shown, with the colormap corresponding to the reduced chi-square. The bottom panels show the one-dimensional posteriors for the two parameters. Blue, red and black lines, correspond to the supernova data, the Hubble parameter data, and their combination, respectively.}
    \label{fig04}
  \end{center}
\end{figure*}

It is also worth briefly considering the case with $r=0$, in which we have the analytic solutions $f(z)=(1+z)$ and $g(z)=(1+z)^2$. Moreover, from Eq. (\ref{middle1}) we can write
\be
1+w_0=-\frac{2}{3}n\,;
\ee
in this case the definition of $\Omega_A$ in Eq. (\ref{defoma}) does not apply, but we define instead $\Omega_A=\kappa^2V_0^2/H_0^2$. As before, in a spatially flat universe this parameter does not explicitly appear in the Friedmann equation, which becomes
\be
E^2=\Omega_m(1+z)^3+(1-\Omega_m)(1+z)^{-2n}\,,
\ee
which recovers $\Lambda$CDM for $n=0$. (Evidently, identical equations would apply for the $m$ case.) In agreement with the discussion in the previous section, we also see that canonical and phantom equations of state obtain for negative and positive values of $n$, respectively. Figure \ref{fig04} shows the constraints on this case. As expected, model parameters consistent with $\Lambda$CDM are preferred by the data. Specifically we find
\bq
\Omega_m&=&0.27\pm0.02\\
n&=&-0.12\pm0.09\,.
\eq
These are consistent, at two standard deviations, with a cosmological constant, and again the preferred value of the matter density fully agrees with the one for the $w_0$CDM model with the same data.

A final particular case is $w_0=-1$, which corresponds to the case where the triad mimics a cosmological constant today, but is allowed to behave differently at non-zero redshifts. Here the choice of potential becomes important, and the results for the two potentials defined in Eqs. (\ref{potential1},\ref{potential2}) must therefore be discussed separately.

\begin{figure*}
  \begin{center}
    \includegraphics[width=0.68\columnwidth]{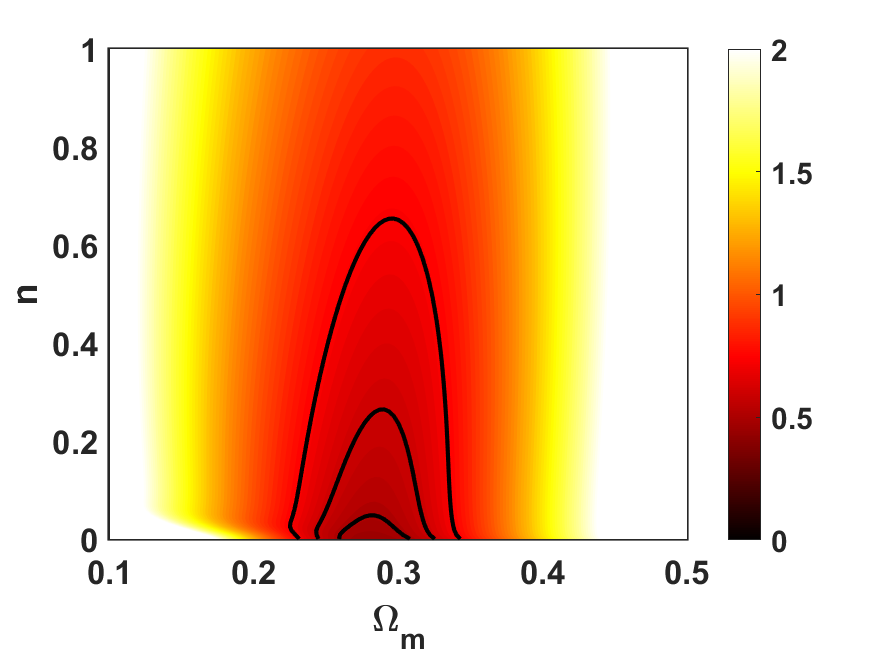}
    \includegraphics[width=0.68\columnwidth]{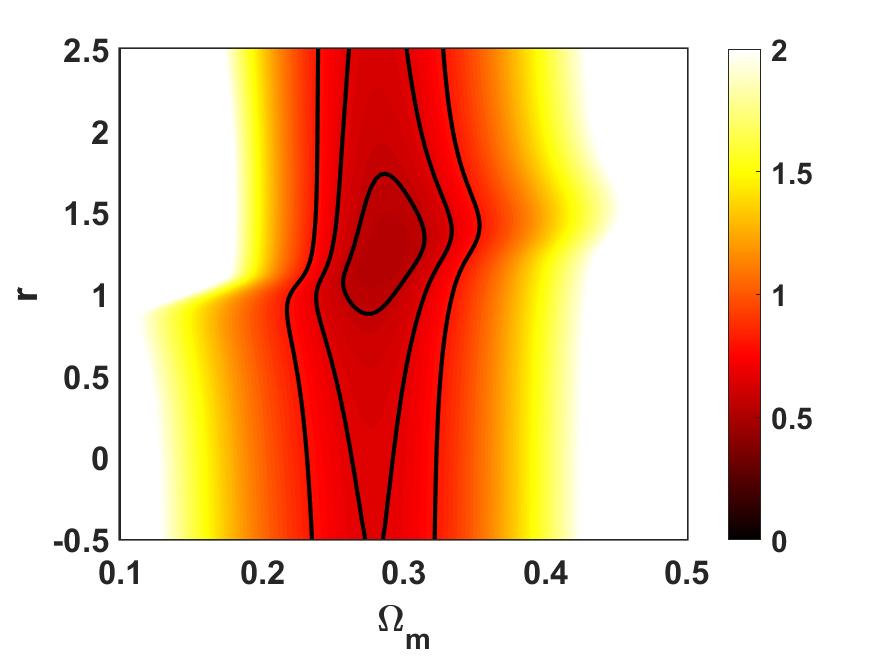}
    \includegraphics[width=0.68\columnwidth]{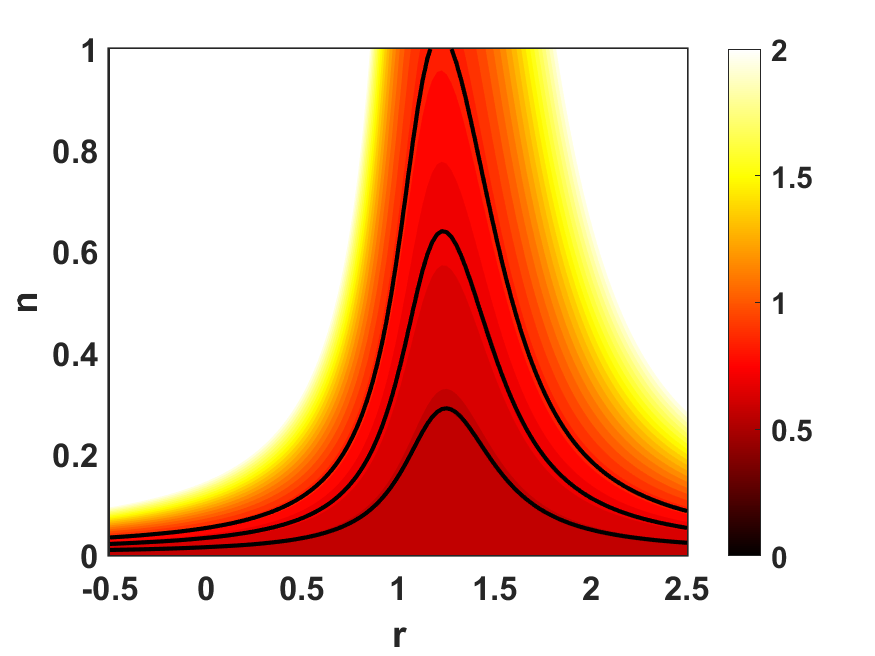}
    \includegraphics[width=0.68\columnwidth]{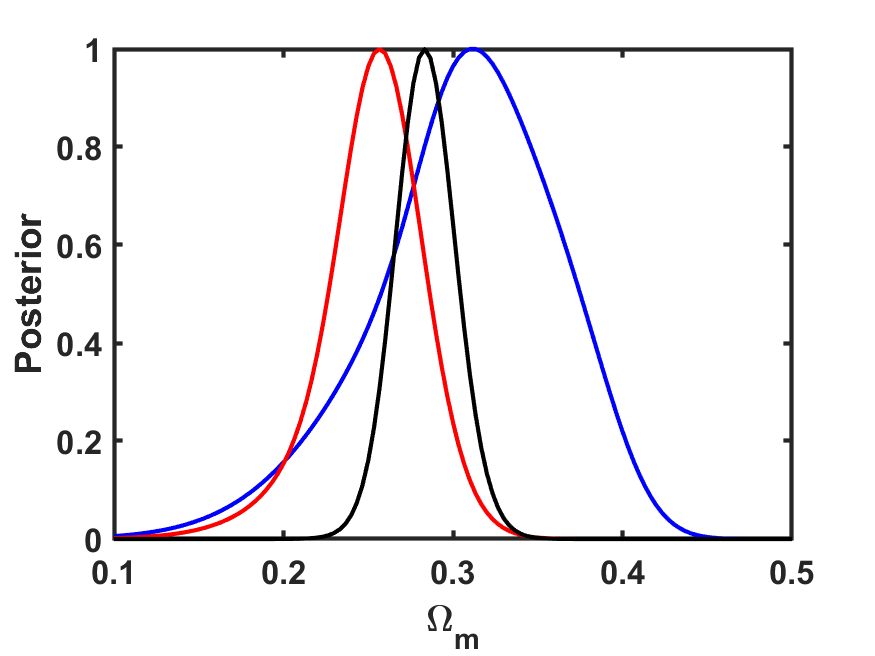}
    \includegraphics[width=0.68\columnwidth]{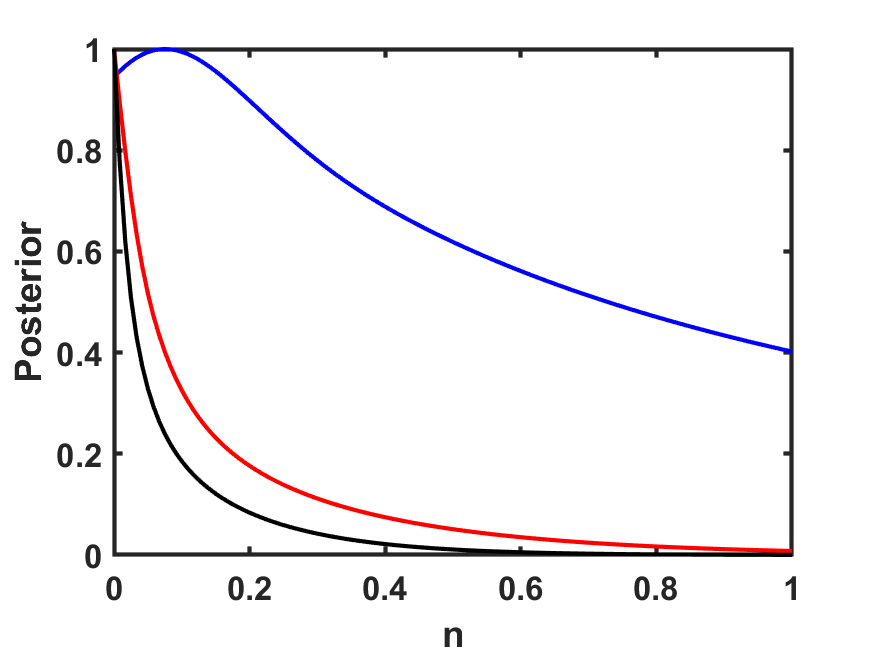}
    \includegraphics[width=0.68\columnwidth]{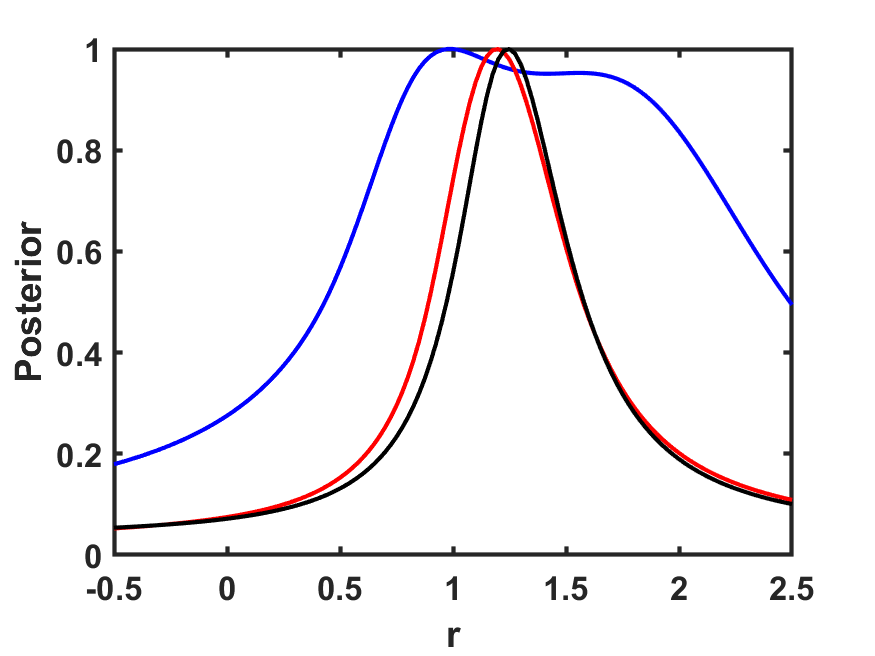}
    \caption{Constraints on the $w_0=-1$ triad model with a power-law potential. The top row panels depict the constraints on the relevant two-dimensional planes: one, two and three sigma confidence levels are shown in black, with the colormap corresponding to the reduced chi-square. The bottom row panels show the one-dimensional posteriors for each parameter. Blue, red and black lines, correspond to the supernova data, the Hubble parameter data, and their combination, respectively.}
    \label{fig05}
  \end{center}
\end{figure*}
\begin{figure*}
  \begin{center}
    \includegraphics[width=0.68\columnwidth]{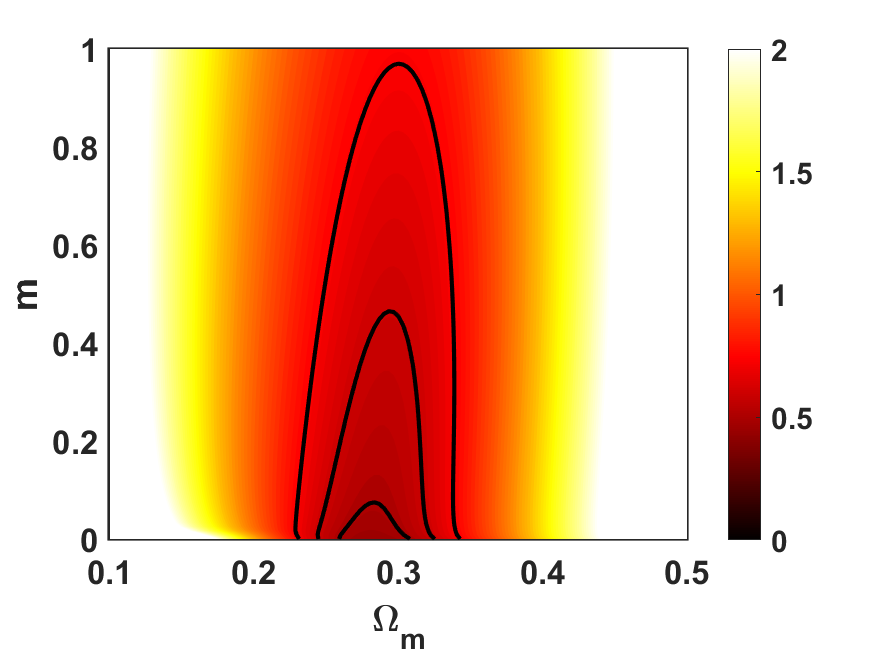}
    \includegraphics[width=0.68\columnwidth]{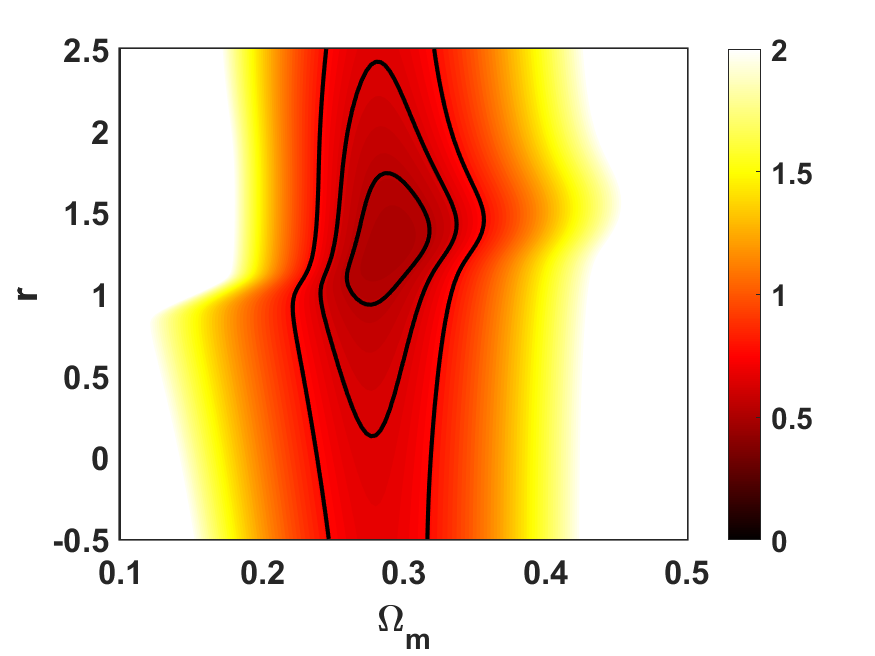}
    \includegraphics[width=0.68\columnwidth]{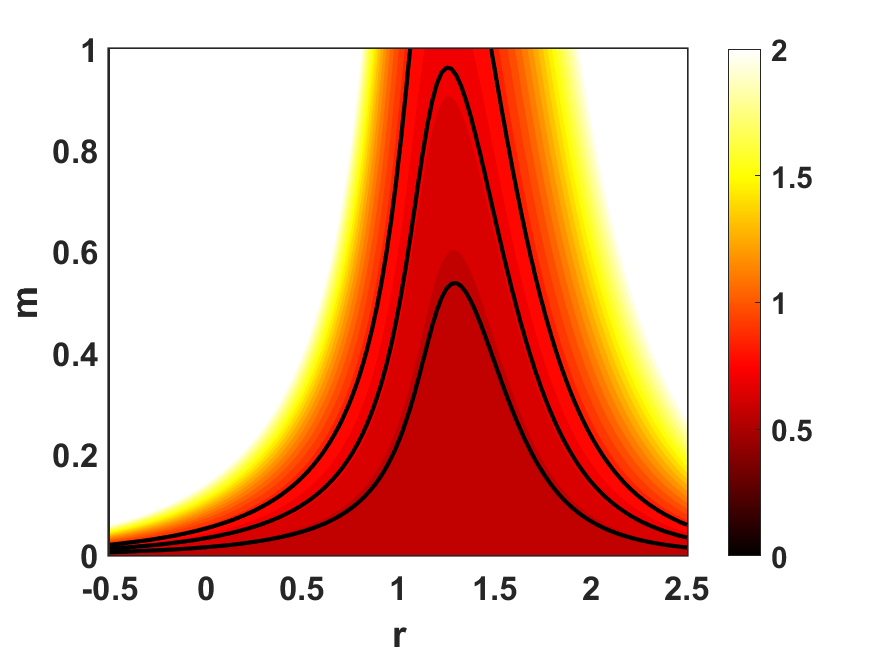}
    \includegraphics[width=0.68\columnwidth]{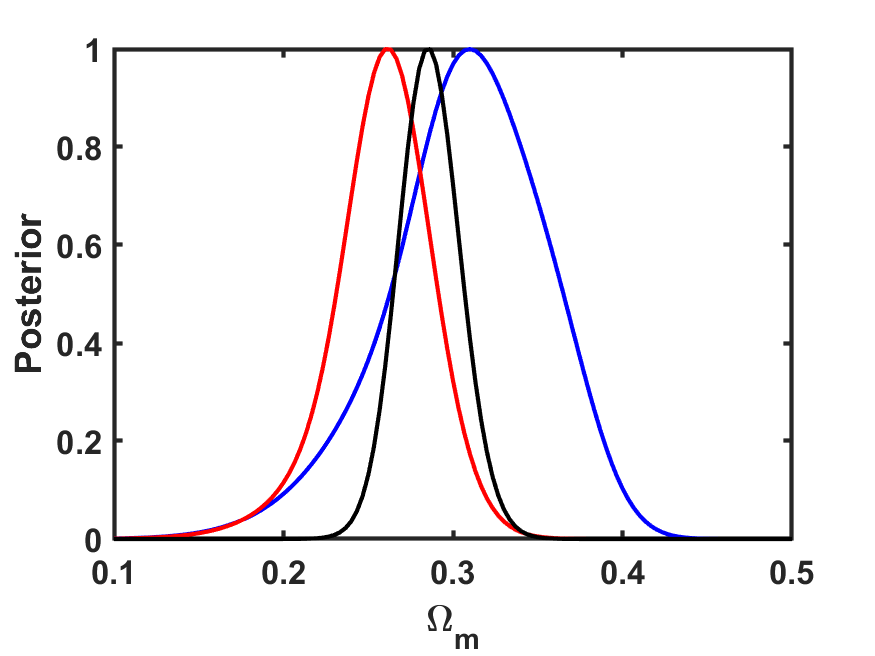}
    \includegraphics[width=0.68\columnwidth]{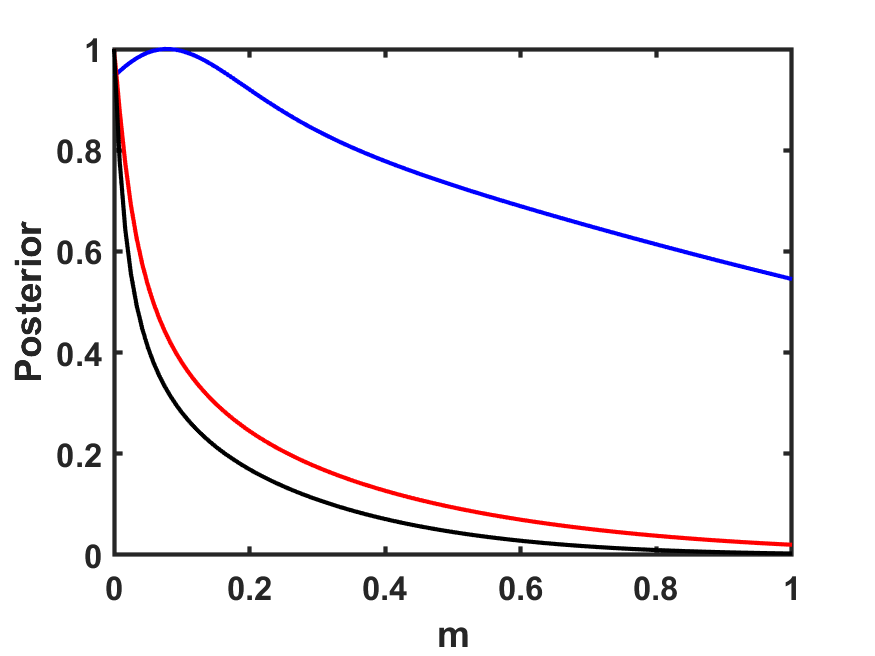}
    \includegraphics[width=0.68\columnwidth]{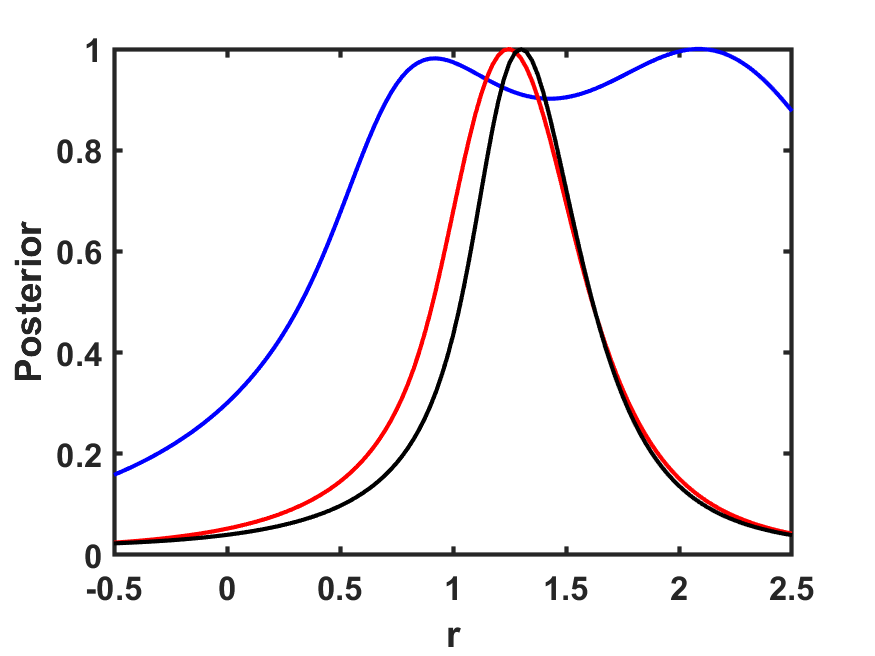}
    \caption{Same as Fig. \ref{fig05}, for the exponential potential.}
    \label{fig06}
  \end{center}
\end{figure*}

For the power law potential, the Friedmann and Proca equations are given by setting $w_0=-1$ in Eqs. (\ref{base0},\ref{bv1},\ref{friedv1}). We restrict ourselves to $n\ge0$, since negative values can easily lead to manifestly unphysical negative values of $E^2$. The results of the analysis are depicted in Fig. \ref{fig05}. One sees that the matter density is well constrained, not being significantly correlated with the other model parameters. These additional parameters, $n$ and $r$ do show some correlation, but can nevertheless be constrained. Specifically, we find the constraints
\bq
\Omega_m&=&0.28\pm0.02\\
n&<&0.14\\
r&=&1.25\pm0.25\,,
\eq
where for the potential slope we report a two-sigma upper limit. Note that the decaying solution $A\propto 1/a$, which would correspond to $r=0$, is clearly not preferred by the data. Instead, a value consistent with $r=1$, which corresponds to ${\dot f_0}=0$, is the statistically preferred one.

For the exponential potential, the analogous evolution equations are the particular $w_0=-1$ cases of Eqs. (\ref{base0},\ref{bv2},\ref{friedv2}), and we restrict ourselves to $m\ge0$ for the same reason. Figure \ref{fig06} shows the results in this case, for which we find the constraints
\bq
\Omega_m&=&0.29\pm0.02\\
m&<&0.25\\
r&=&1.30\pm0.25\,.
\eq
The two results are quite similar, since the data constrains the potentials to be nearly constant, in which limit there is no substantive difference between them. The main salient point is that $n$ is substantially more constrained than $m$.

\begin{figure*}
  \begin{center}
    \includegraphics[width=0.68\columnwidth]{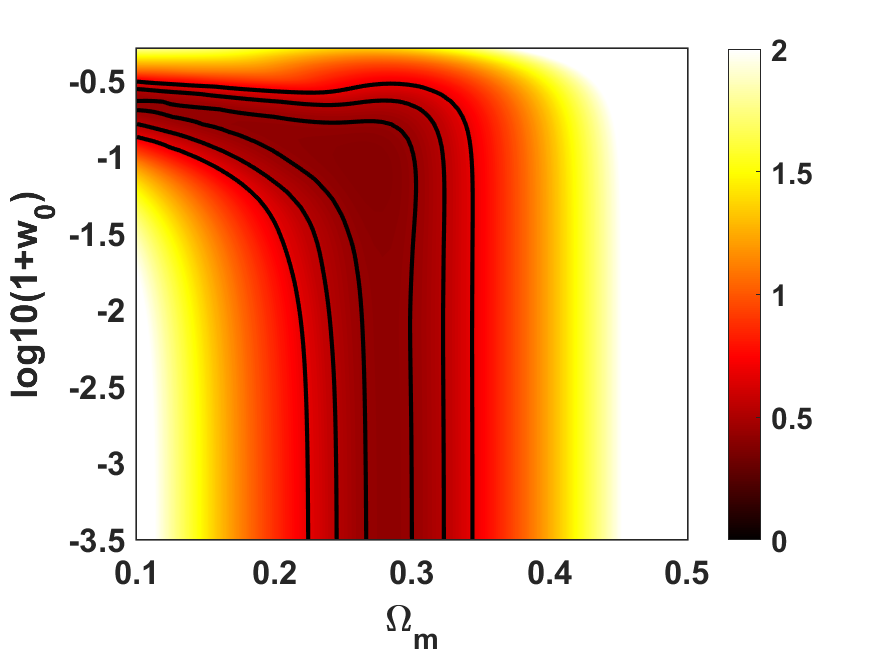}
    \includegraphics[width=0.68\columnwidth]{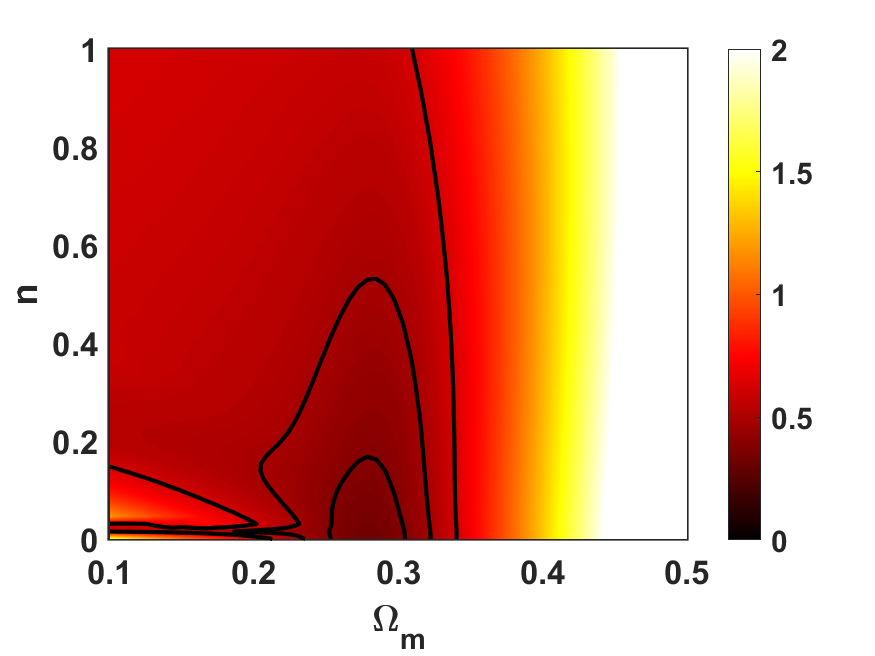}
    \includegraphics[width=0.68\columnwidth]{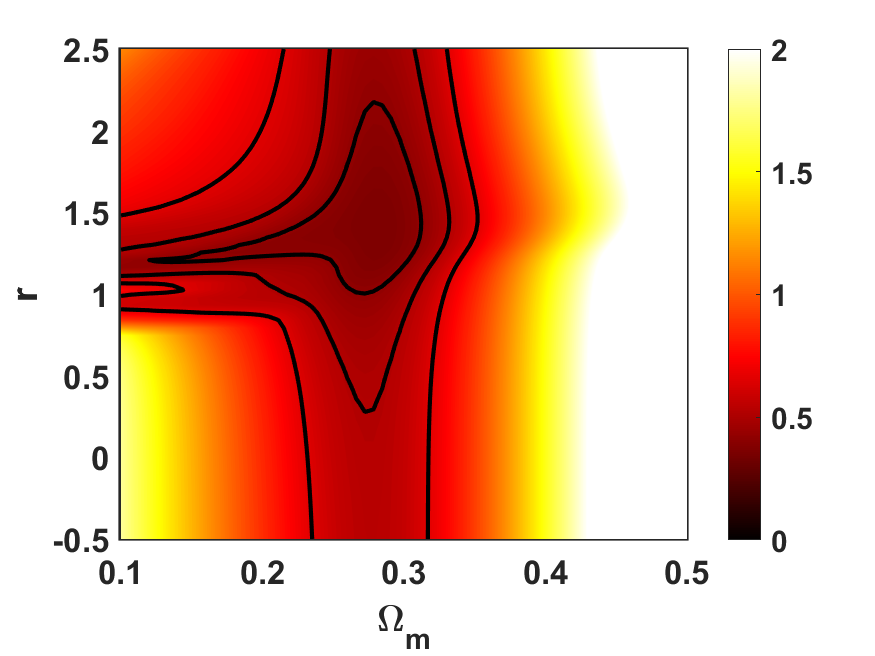}
    \includegraphics[width=0.68\columnwidth]{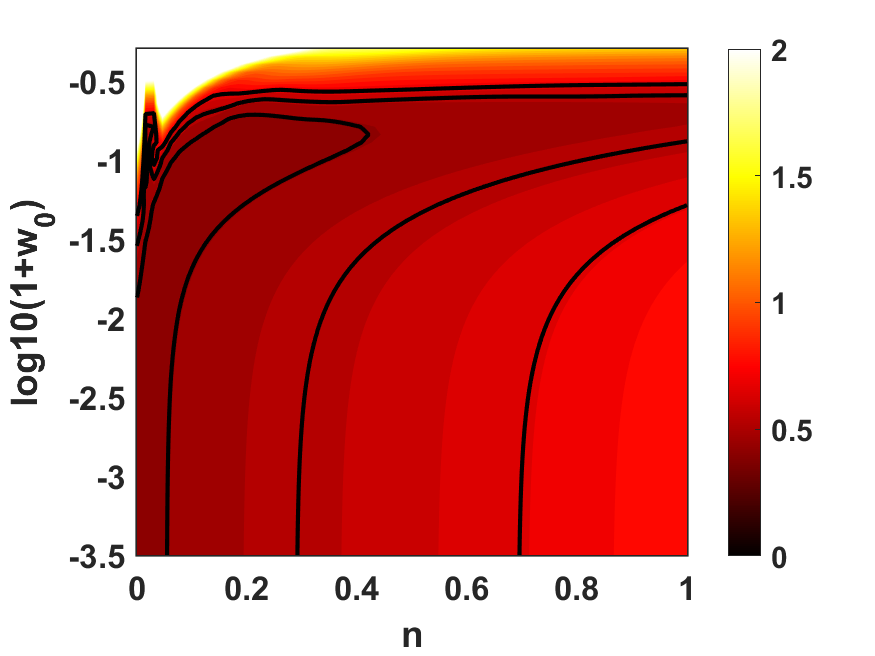}
    \includegraphics[width=0.68\columnwidth]{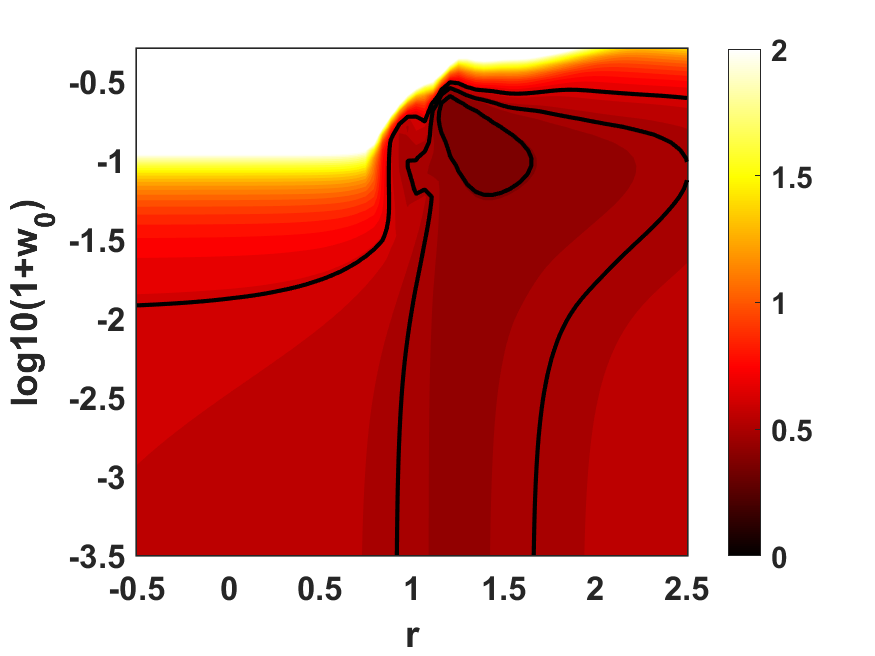}
    \includegraphics[width=0.68\columnwidth]{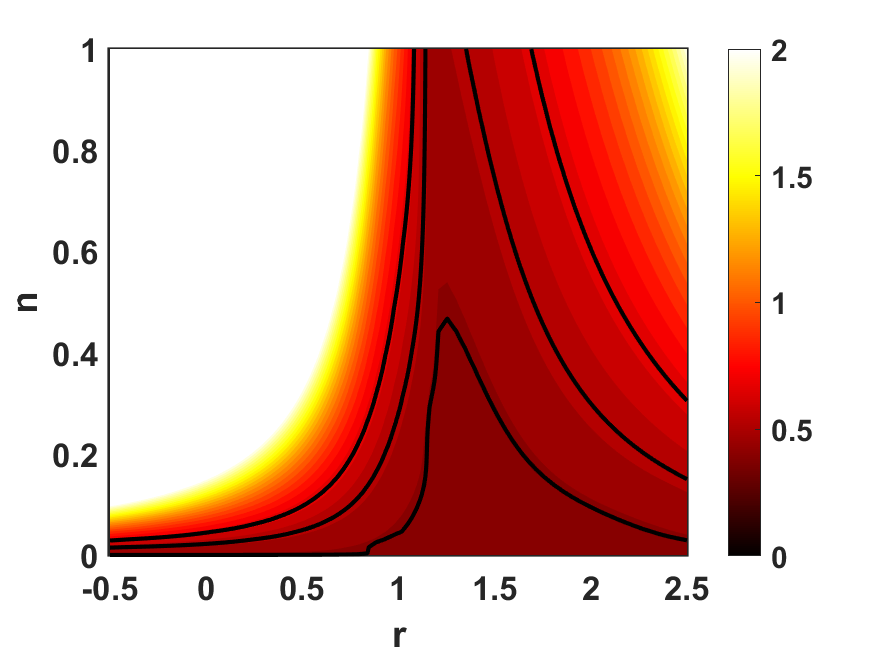}
    \includegraphics[width=0.51\columnwidth]{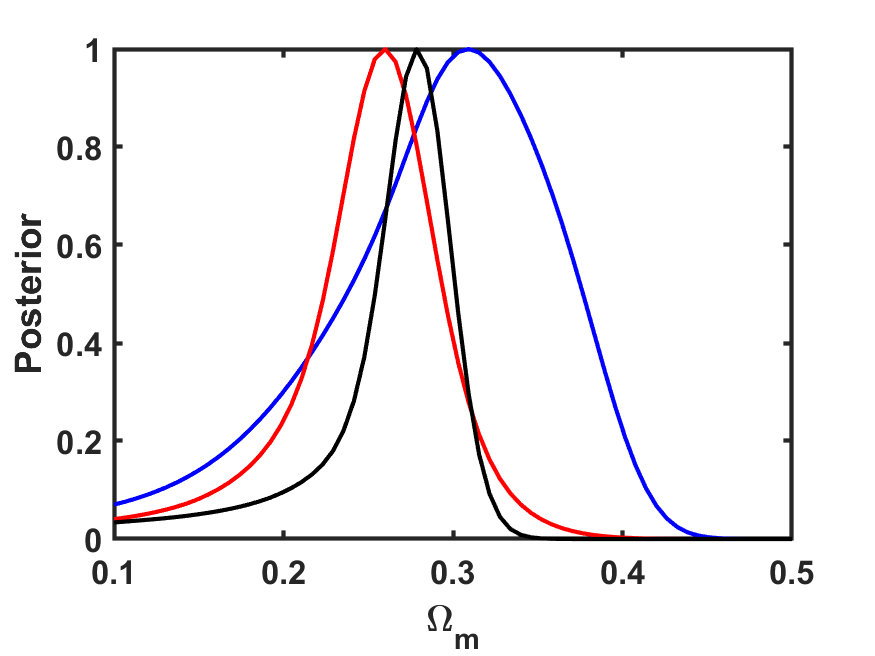}
    \includegraphics[width=0.51\columnwidth]{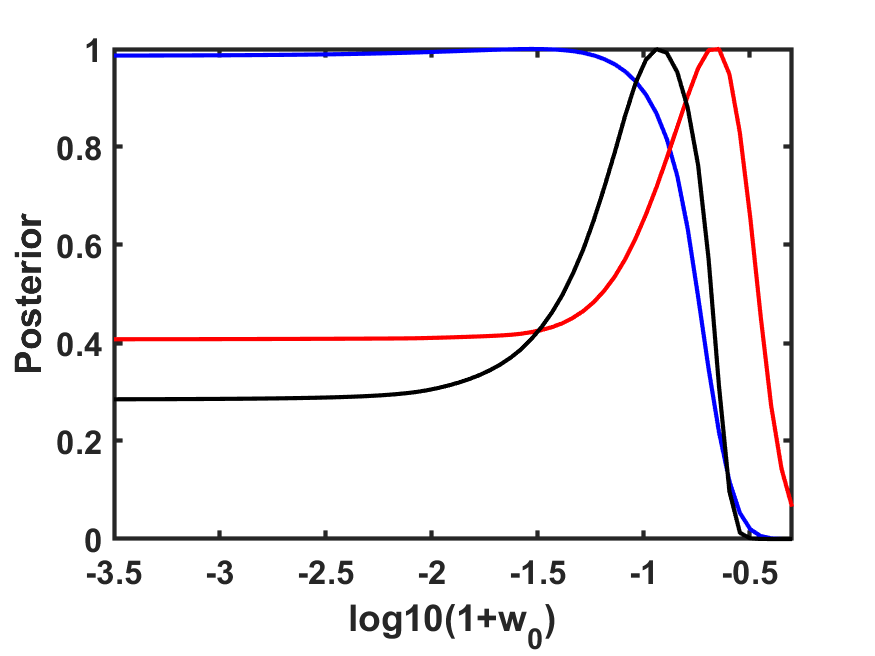}
    \includegraphics[width=0.51\columnwidth]{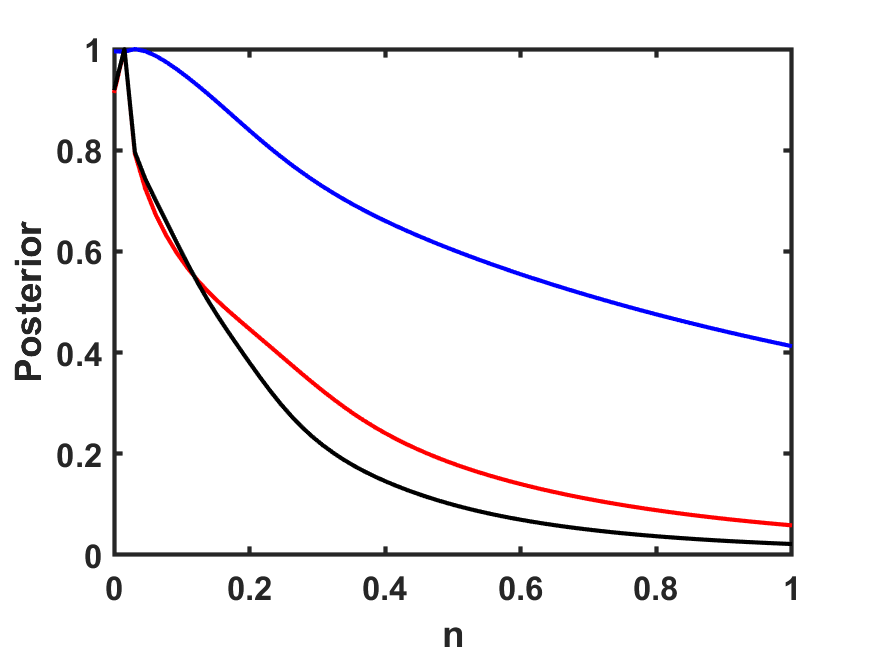}
    \includegraphics[width=0.51\columnwidth]{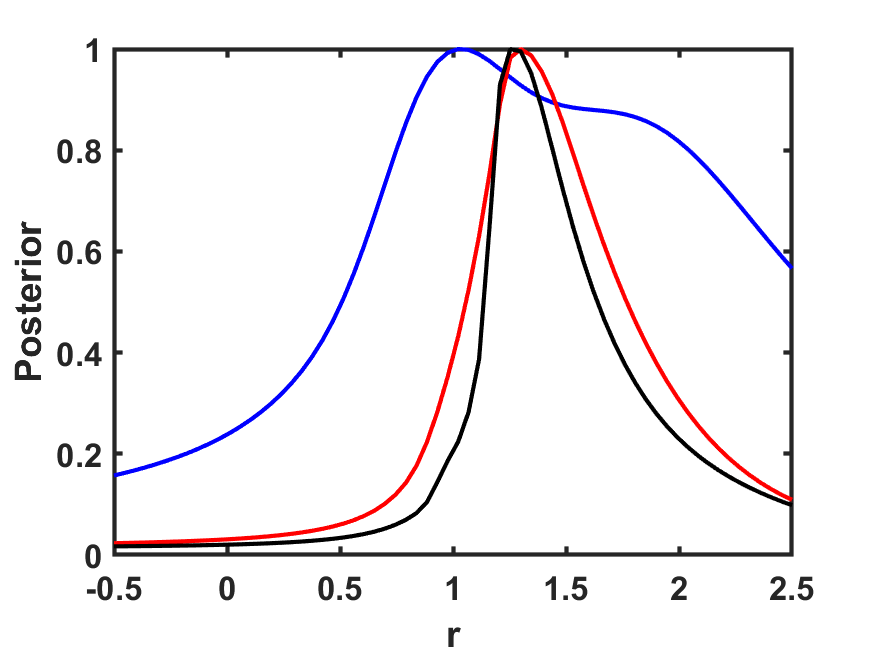}
    \caption{Constraints on the full triad model with a power-law potential and a logarithmic prior on the dark energy equation of state. The top and middle row panels depict the constraints on the relevant two-dimensional planes: one, two and three sigma confidence levels are shown in black, with the colormap corresponding to the reduced chi-square. The bottom panels show the one-dimensional posteriors for each parameter. Blue, red and black lines, correspond to the supernova data, the Hubble parameter data, and their combination, respectively.}
    \label{fig07}
  \end{center}
\end{figure*}
\begin{figure*}
  \begin{center}
    \includegraphics[width=0.68\columnwidth]{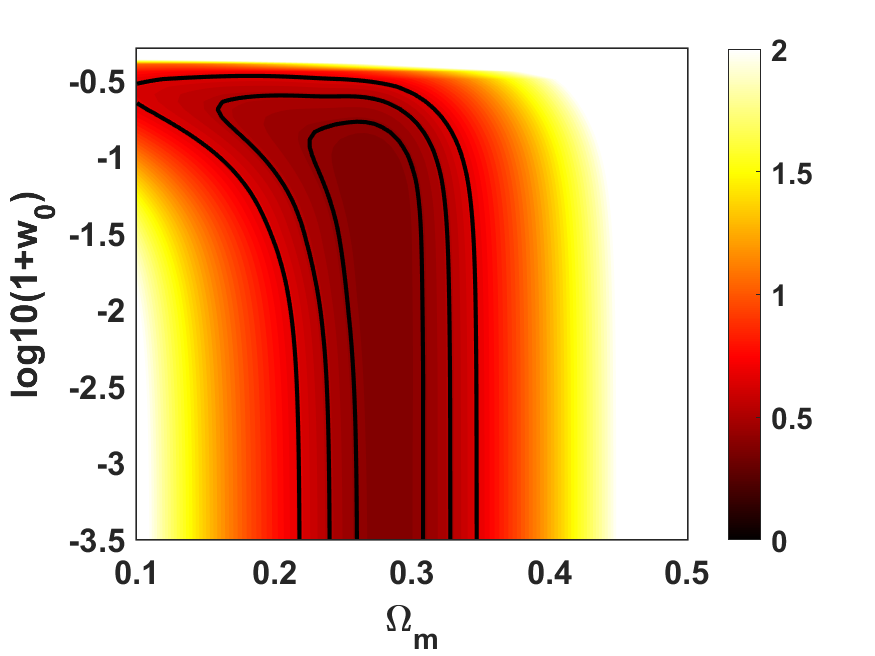}
    \includegraphics[width=0.68\columnwidth]{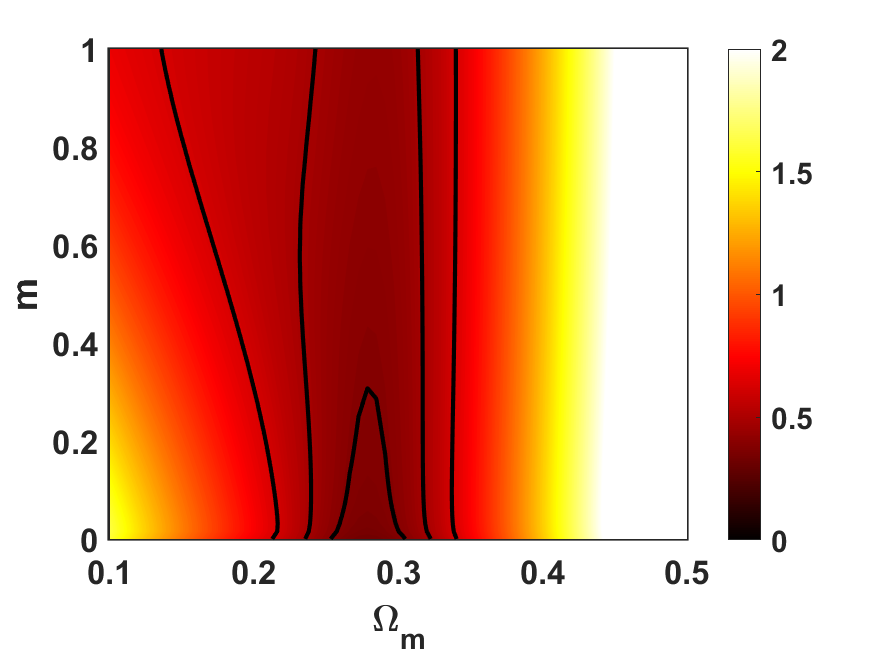}
    \includegraphics[width=0.68\columnwidth]{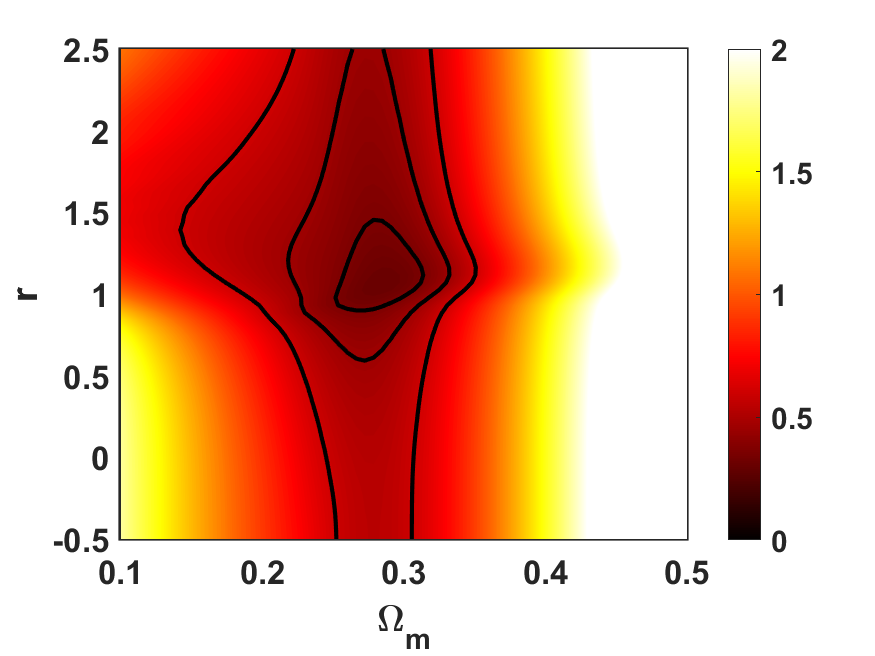}
    \includegraphics[width=0.68\columnwidth]{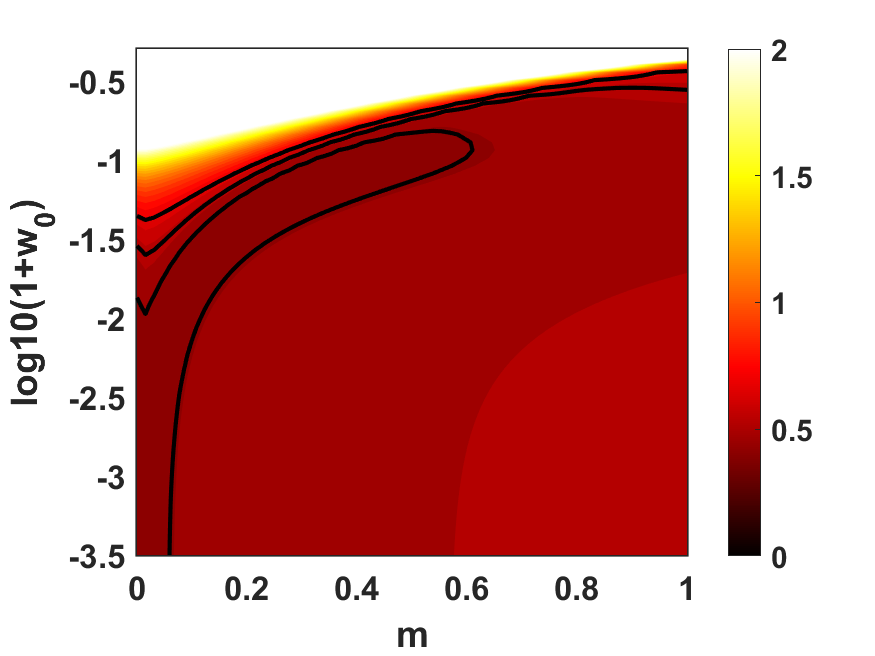}
    \includegraphics[width=0.68\columnwidth]{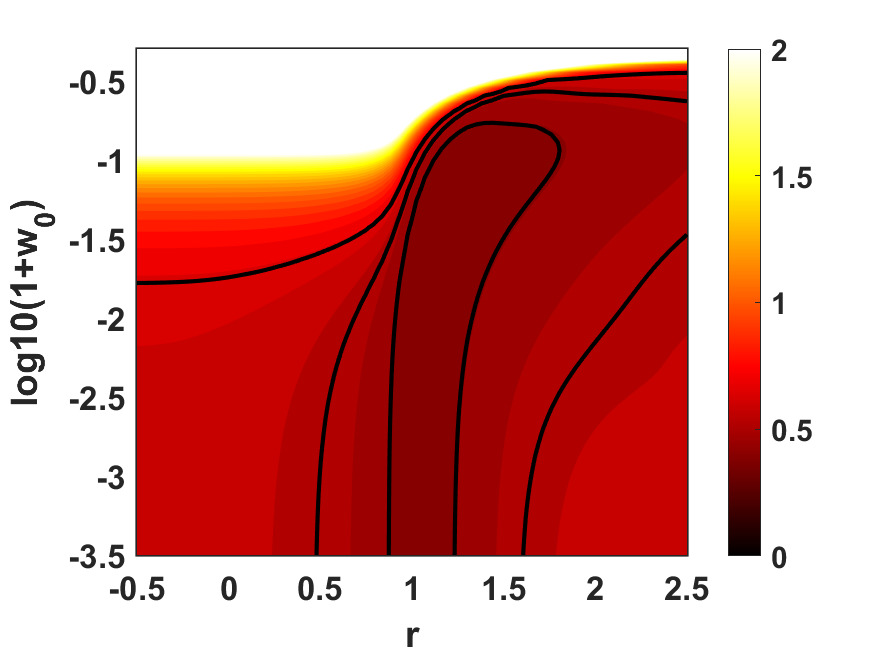}
    \includegraphics[width=0.68\columnwidth]{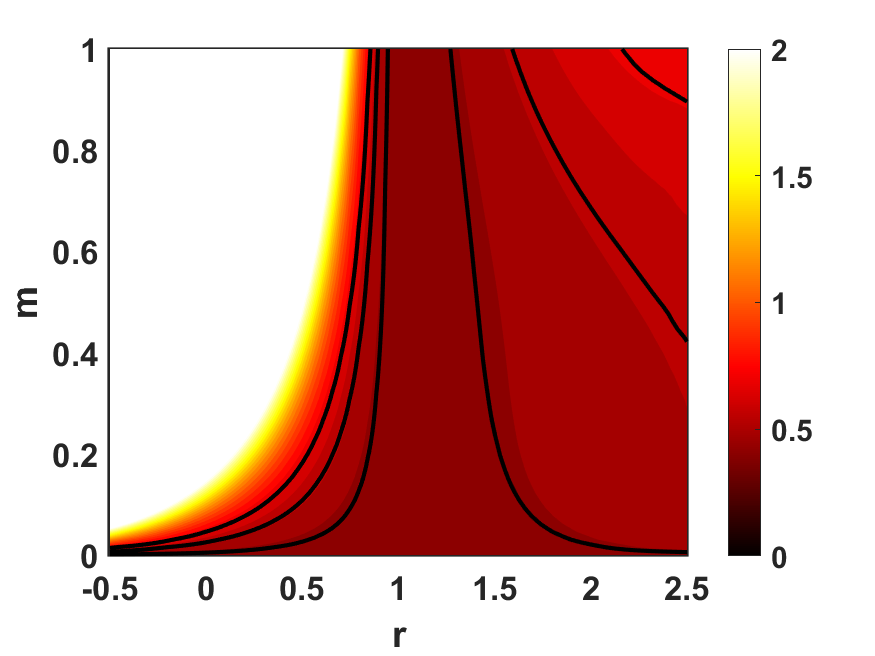}
    \includegraphics[width=0.51\columnwidth]{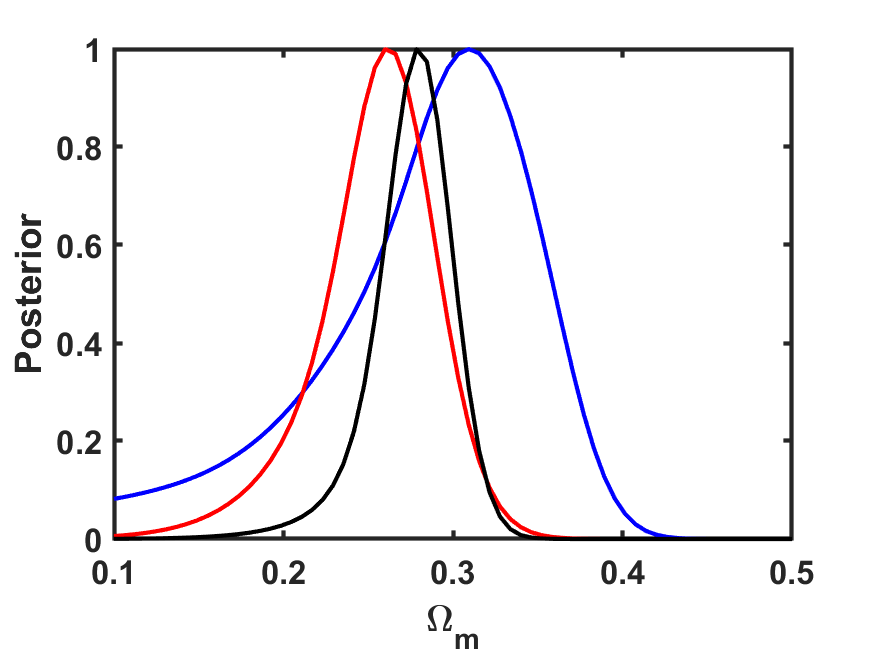}
    \includegraphics[width=0.51\columnwidth]{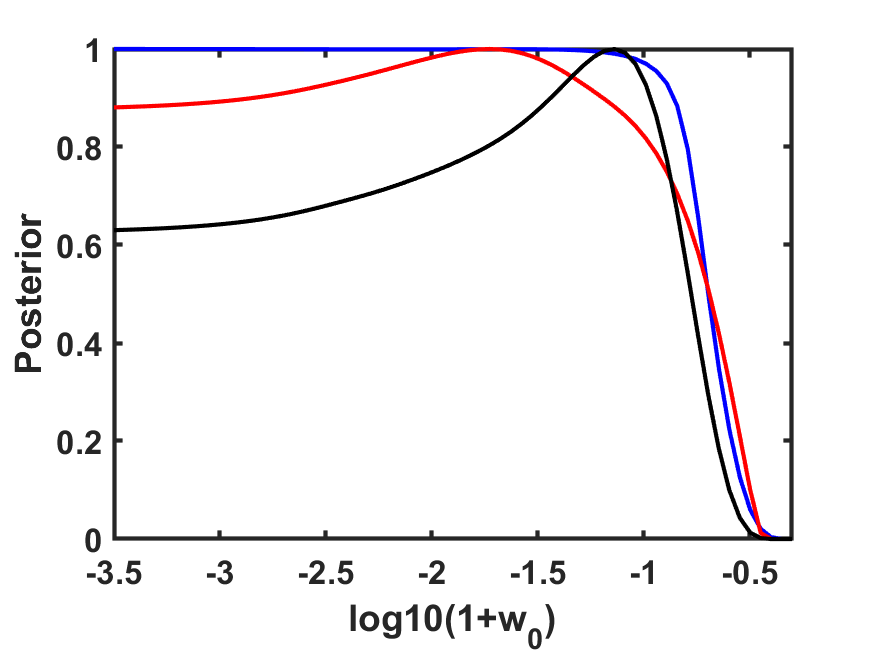}
    \includegraphics[width=0.51\columnwidth]{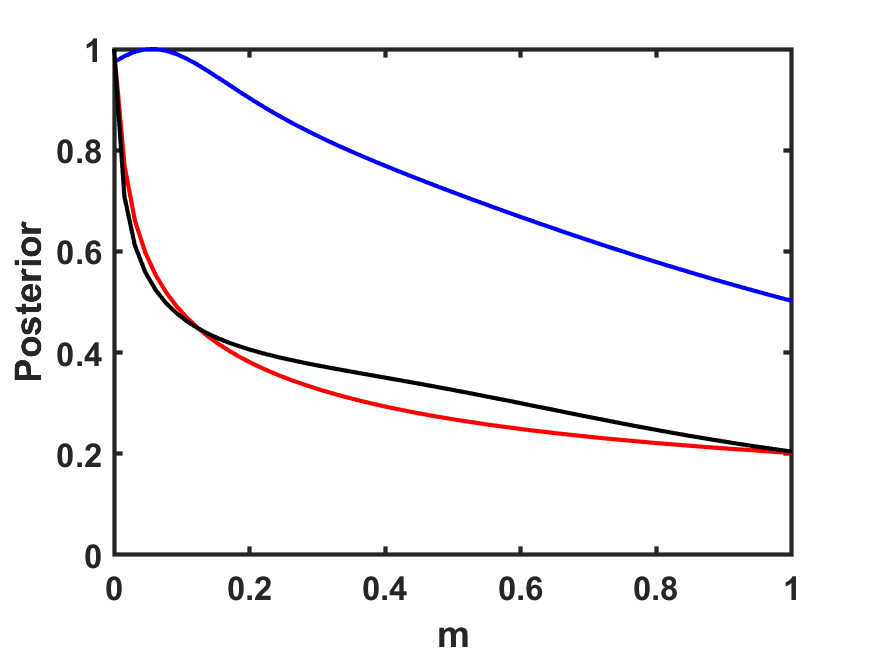}
    \includegraphics[width=0.51\columnwidth]{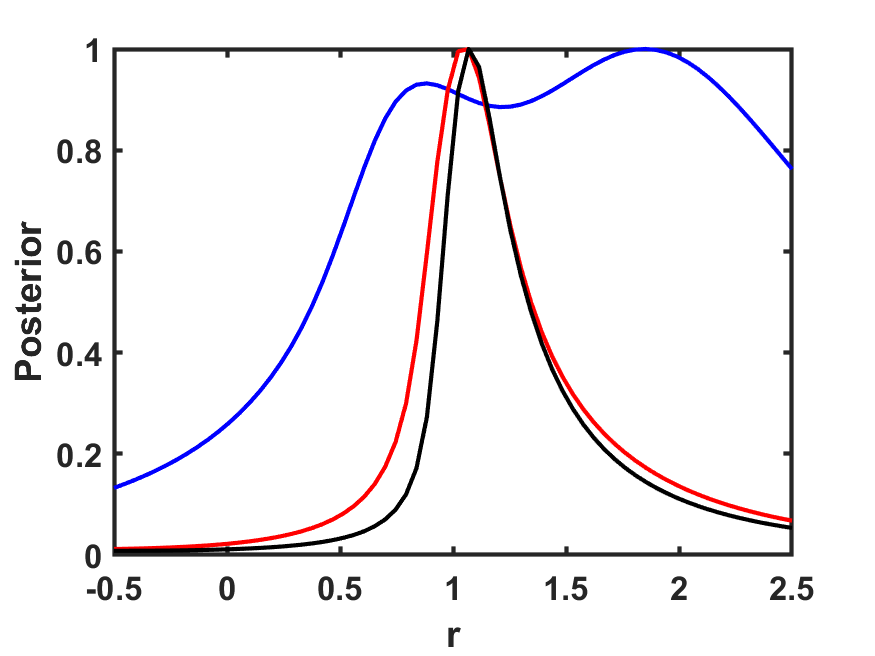}
    \caption{Same as Fig. \ref{fig07}, for the exponential potential.}
    \label{fig08}
  \end{center}
\end{figure*}

\section{\label{result2}Constraints for generic triads}

We now separately present the results for the two choices of potential for the full triad model. We retain the physically safer assumptions of $n\ge0$ (and similarly $m\ge0$), as well as $w_0\ge-1$, for the reasons already explained. We will also explore the impact of the choices of priors, specifically considering two alternative assumptions on the present-day equation a state: a uniform prior on $w_0$, or a logarithmic prior---i.e., a uniform prior on $\log_{10}(1+w_0)$

Figures \ref{fig07} and \ref{fig08} show the results for the two potentials, for the latter choice of priors. For simplicity we do not show analogous plots for the alternative choice of priors, but we do report, in Table \ref{table1}, the derived posterior constraints on the four model parameters, for the four possible combinations of triad potential and equation of state prior.

\begin{table*}
\centering
\caption{Posterior one-sigma constraints, or two-sigma upper limits, for the parameters of the general triad model, with the choices of potential and dark energy equation of state prior discussed in the main text.}
\begin{tabular}{| c | c | c | c | c | c |}
\hline
Potential & Prior & Matter density & Equation of state & Potential slope  & Triad speed \\
\hline
Power law & Uniform & $\Omega_m=0.27\pm0.02$ & $w_0=-0.88^{+0.07}_{-0.08}$ & $n=0.17^{-0.12}_{+0.15}$ & $r=1.21_{-0.04}^{+0.23}$ \\
Power law & Logarithmic & $\Omega_m=0.28\pm0.02$ & $\log_{10}(1+w_0)<-0.61$ & $n<0.42$ & $r=1.23_{-0.09}^{+0.21}$ \\
\hline
Exponential & Uniform & $\Omega_m=0.27\pm0.02$ & $w_0=-0.93^{+0.07}_{-0.08}$ & Unconstrained & $r=1.25_{-0.18}^{+0.43}$ \\
Exponential & Logarithmic & $\Omega_m=0.28\pm0.02$ & $\log_{10}(1+w_0)<-0.62$ & Unconstrained & $r=1.07_{-0.11}^{+0.20}$ \\
\hline
\end{tabular}
\label{table1}
\end{table*}

Overall, our results are consistent with those for the particular cases, reported in the previous section. The first noteworthy point is that despite the extended parameter space the matter density is still tightly constrained, and its preferred value remains the same as before, regardless of the choice of potential and prior. The other three model parameters are more significantly correlated with one another, and their constraints are impacted by these correlations.

The upper limits on the potential slope are significantly relaxed, by a factor of 3 in the case of $n$ for the power-law model. For the power-law model and a uniform $w_0$ prior the one-sigma posterior is non-zero, but at two sigma one gets an upper limit; the latter also obtains for a logarithmic prior. For the exponential model, the result in the previous section that $m$ is more weakly constrained still holds, and in fact in this four-dimensional parameter space $m$ is unconstrained at two standard deviations. As for the triad speed $r$, it is still the case that the present-day field speed is constrained to be small, but a positive speed (${\dot f}_0$) is preferred. The main salient point regarding the posterior likelihood for $r$ is that in the previous section, where the parameter spaces under study were simpler, it was approximately Gaussian near the peak, but this is clearly no longer the case in the full parameter space.

Last but not least, for the triad's equation of state, with uniform priors we again get values which are consistent with a cosmological constant, at one and two standard deviations for the exponential and power-law potentials, respectively. As for the logarithmic prior, in the case of the power-law potential there is a one-sigma (therefore not statistically significant) preference for $\log_{10}(1+w_0)=-0.94_{-0.35}^{+0.24}$; note that $\log_{10}(1+w_0)=-0.61$ corresponds to $w_0=-0.885$, in excellent agreement with the best-fit $w_0$ value in the uniform prior case. Still, at the two-sigma level one finds upper limits for the two potentials, which moreover are very similar to each other: $\log_{10}(1+w_0)<-0.61$ corresponds to $w_0<-0.75$, still a tight constraint considering the degeneracies with other model parameters.

\section{\label{conc}Conclusions}

We have presented quantitative constraints on a class of phenomenological models, previously proposed by Armend\'ariz-Pic\'on and known as the cosmic triad, in which the recent accelerating phase of the universe is due to a suitable combination of vector fields and not to the more commonly invoked scalar fields. As shown in \cite{Paper1}, three mutually orthogonal vector fields can lead to some accelerating universe solutions while preserving large-scale homogeneity and isotropy. Exploring this scenario is therefore a stress test of the canonical $\Lambda$CDM model and its scalar field based alternatives. These models (with their specific choices of potentials) have been previously proposed by other authors but not subject to a detailed quantitative comparison with observations. Our work provides this analysis.

We have restricted ourselves to background cosmology observables, as in these vector-based models linear cosmological perturbation theory includes two non-trivial features: there are anisotropic stresses, and (more significantly) the usual decoupling of scalar, vector and tensor perturbations from each other does not apply \cite{Paper1}. Addressing these issues is beyond the scope of this work, though we note recent attempts to address them \cite{Koyama,Chase}. We have also restricted ourselves to flat universes, setting the spatial curvature parameter to zero, in agreement with most contemporary cosmological observations.

As shown in several relevant limits, the triad model is a parametric extension of the canonical $\Lambda$CDM model, and our analysis demonstrates that even low-redshift background cosmology data suffices to tightly constrain the model to be close to the canonical behavior. Specifically, the preferred value of the matter density always coincides with its standard best-fit value for the same cosmological datasets, and the correspondence is also quite close for the present-day value of the dark energy equation of state.

Considering two possible choices of the triad's potential, each including a single free parameter (effectively a potential slope), suggested in different pervious works \cite{Paper1,Paper3}, we find only a weak dependence on this choice: to a first approximation, the potential is constrained to be nearly flat, which naturally corresponds to the $\Lambda$CDM limit. Unsurprisingly, the constraints are mildly dependent on whether phantom values of the present-day dark energy equation of state, $w_0$, are allowed. Making the physically safer assumption that they are not, which we did for most of our analysis, the obtained constraints do not significantly depend on whether a uniform or logarithmic prior is chosen for $w_0$.

It is interesting to contrast the triad model with the dyad model discussed in \cite{Paper2a,Paper2b}, which we briefly addressed in the appendix. The dyad model does not include a potential and is not a parametric extension of flat $\Lambda$CDM. One consequence which stems from these properties, already pointed out in the original articles (which we have confirmed), is that under a spatial flatness assumption the model is ruled out, and a good fit to the data would instead require a closed universe.

Overall, our analysis shows that, while one may suspect that $\Lambda$CDM is not a fully correct description of cosmological evolution (and, in particular, of its recent acceleration phase and the underlying physical mechanism), but only a simple approximation thereof, it is nevertheless a convenient and reasonably accurate one. From a model-building perspective, one or more scalar fields are certainly the most convenient alternative to a cosmological constant, but they are not the only one: the triad model shows that vector field based models can lead to very similar phenomenology. Ultimately, cosmology is a data-driven science, and distinguishing between different classes of models all of which behave, on the largest scales, very similarly to a cosmological constant, may be unfeasible if one relies only on traditional observables. Instead, one may need precision tests of cornerstone cosmological principles, e.g. of the Einstein Equivalence Principle and of the universality of physical laws \cite{Will,ROPP,MICROSCOPE,Peebles}.

\begin{acknowledgments}
This work was financed by Portuguese funds through FCT (Funda\c c\~ao para a Ci\^encia e a Tecnologia) in the framework of the project 2022.04048.PTDC (Phi in the Sky, DOI 10.54499/2022.04048.PTDC). CJM also acknowledges FCT and POCH/FSE (EC) support through Investigador FCT Contract 2021.01214.CEECIND/CP1658/CT0001 (DOI 10.54499/2021.01214.CEECIND/CP1658/CT0001). 

The project that led to this work was started during AstroCamp 2023. CCC and ALG were partially supported by AstroCamp funds, CCC was also partially supported by Ci\^encia
Viva OCJF funds.
\end{acknowledgments}

\appendix
\section{\label{app1}The dyad model}

Here we present a brief discussion of the dyad model \cite{Paper2a,Paper2b}, comparing it with the triad model which was considered in detail in the main text and relying on a similar analysis methodology. In contrast to the triad model, \cite{Paper2a} build a model relying on two vector fields and suitable derivatives thereof (or alternatively, a gauge vector field with a gauge-fixing term) coupled to the Ricci tensor, but including no potential. This is a physically pathological model, encountering a future singularity in a finite time.

The energy density and pressure of the cosmic dyad and its Proca-like equation take the following forms, respectively
\bq
\rho_A&=&\frac{3}{2}H^2A^2+3HA{\dot A}-\frac{1}{2}{\dot A}^2\\
p_A&=&-3\left(\frac{5}{2}H^2+\frac{4}{3}{\dot H}\right)A^2+HA{\dot A}-\frac{3}{2}{\dot A}^2\,,
\eq
\be\label{proca2}
{\ddot A}+3H{\dot A}-3\left(H^2+\frac{\ddot a}{a}\right)A=0\,;
\ee
note that for the latter equation the term in brackets we can alternatively write $H^2 + \Ddot{a}/a = 2H^2 + \Dot{H}$, which is proportional to the Ricci scalar. In this case there is no $\Lambda$CDM limit, as already noticed in \cite{Paper2b}, who also point out that in this model a closed universe would be necessary.

The vector field scaling solutions analogous to those of Eqs. (\ref{conformal1},\ref{conformal2}) can generically be written
\be
\alpha_A=\frac{1-3\beta\pm\sqrt{33\beta^2-18\beta+1}}{2}\,;
\ee
for these solution to be real the term inside the square root must be positive, which only occurs outside the range $\beta=(9\pm\sqrt{48})/33\sim[0.06,0.48]$. For the radiation era the scaling solutions are still the constant one and the one given by Eq. (\ref{conformal1}), and the constant solution also has a radiation equation of state, $w_A=1/3$. On the other hand, in the matter era the decaying and growing modes behave approximately as $A\propto t^{-1.46}$ and $A\propto t^{0.46}$ respectively.

\begin{figure}
  \begin{center}
    \includegraphics[width=\columnwidth]{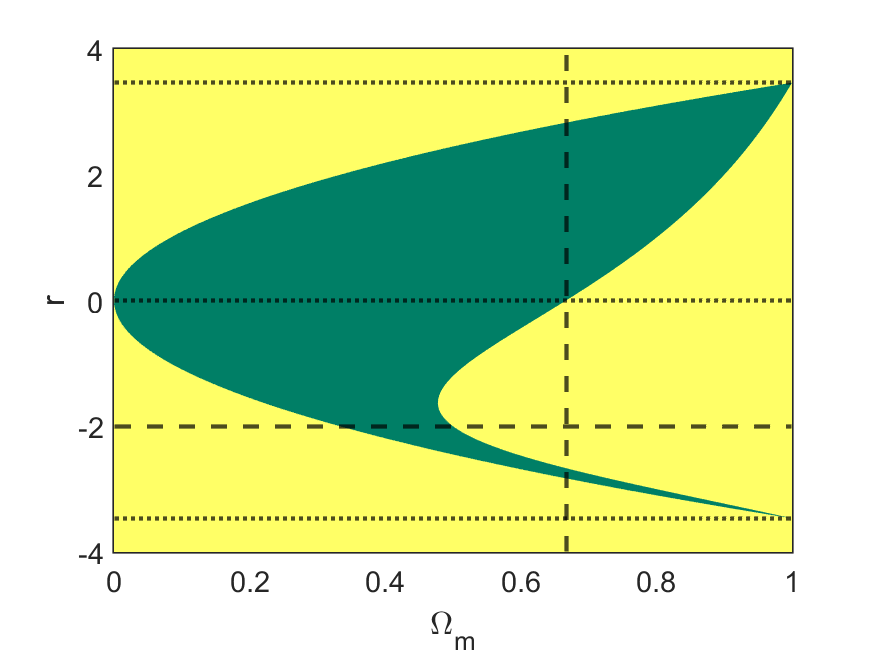}
    \caption{Regions of the $(\Omega_m,r)$ parameter space in which the term in square brackets in Eq.(\ref{taylorc}) has positive and negative values (light and dark colored regions respectively). The dotted lines correspond to $r=0$ and $r=\pm\sqrt{12}$ and the dashed ones identify $r=-2$ and $\Omega_m=2/3$, whose relevance is discussed in the main text.}
    \label{figA}
  \end{center}
\end{figure}

A numerical implementation procedure analogous to that of Sect. \ref{methods} can be applied. In this case the convenient definition is
\be
B=\dot{A}-3HA\,;
\ee
keeping the previous definitions of $f$ and $g$, and using $r\equiv B_0/A_0H_0={\dot f_0}/H_0-3$, the evolution equations for the two functions become
\bq\label{ftwo}
(1+z)\frac{df(z)}{dz}&=&-3f(z)-r\frac{g(z)}{E(z)}\\
(1+z)\frac{dg(z)}{dz}&=&6g(z)+12E(z)\frac{f(z)}{r}\,;\label{gtwo}
\eq
the latter provides a first clue of the different nature of the model, having $r$ in the denominator. With the slightly different definition
\be
\Omega_A = \frac{\kappa^2 B_0^2}{3H_0^2}\,,
\ee
the Friedmann equation can be written
\be
E^2=\Omega_m(1+z)^3+\Omega_A\left(6\frac{f^2E^2}{r^2}-\frac{1}{2}g^2\right)\,,
\ee
or equivalently
\be\label{fred}
E^2=\frac{(r^2-12)\Omega_m(1+z)^3+(1-\Omega_m)r^2g^2}{(r^2-12)+12(1-\Omega_m)f^2}\,;
\ee
here the parameter $r$ does appear explicitly in the Friedmann equation. In this case there is no $\Lambda$CDM limit, but two other limits can be straightforwardly found. For $r\to0$ we have $f=(1+z)^{-3}$ and this becomes
\be
E^2=\frac{\Omega_m(1+z)^9}{(1+z)^6-(1-\Omega_m)}\,,
\ee
while for $r\to\infty$ we have $g=(1+z)^{6}$ and therefore
\be
E^2=\Omega_m(1+z)^3+(1-\Omega_m)(1+z)^{12}\,.
\ee
Moreover, if one Taylor expands Eqs. (\ref{ftwo},\ref{gtwo},\ref{fred}) at zero redshift, one finds
\bq
f&=&1-(3+r)z\\
g&=&1+6\left(1+\frac{2}{r}\right)z\\
E^2&\simeq&1+3z\left[1+(1-\Omega_m)\frac{3r^2+16r+24}{r^2-12\Omega_m}\right]\,.\label{taylorc}
\eq
Figure \ref{figA} identifies the regions of the $(\Omega_m,r)$ parameter space in which the term in square brackets in Eq.(\ref{taylorc}) has positive and negative values; clearly the latter will not fit the data and are even unphysical, which is the case for a substantial part of the parameter space with $r^2\le12$.

Eq.(\ref{taylorc}) can be compared to the analogous expression for flat $\Lambda$CDM,
\be
E^2\simeq 1+3\Omega_m z\,;
\ee
if one demands that the two match, we must satisfy
\be
r=-2\pm\sqrt{3\Omega_m-2}\,,
\ee
but this only has real roots for $\Omega_m\ge2/3$, so there is no real physical match between the two models.

Therefore, with our assumption of a flat universe there's no choice of parameters which provides a good fit to the data. Moreover, a brief exploration also shows that constraints from the two individual datasets are not mutually consistent, e.g. the Hubble parameter data prefers much lower matter densities than the supernova data---a feature which can also be seen in Figure 2 of \cite{Paper2b}. The conclusion is that the only scenario in which which this model might be observationally plausible is if one relaxes the assumption of spatial flatness and allows for the presence of substantial spatial curvature, in agreement with the findings in \cite{Paper2b}.

\bibliography{article}
\end{document}